\documentclass[sn-nature]{sn-jnl}

\usepackage{graphicx}%
\usepackage{multirow}%
\usepackage{amsmath,amssymb,amsfonts}%
\usepackage{amsthm}%
\usepackage{mathrsfs}%
\usepackage[title]{appendix}%
\usepackage{xcolor}%
\usepackage{textcomp}%
\usepackage{manyfoot}%
\usepackage{algorithm}%
\usepackage{algorithmicx}%
\usepackage{algpseudocode}%
\usepackage{listings}%

\usepackage{bm}
\usepackage{multirow}
\usepackage{subcaption}
\usepackage{booktabs}
\usepackage{hyperref}
\usepackage{soul}
\usepackage{float}
\usepackage{tabularx}
\usepackage{cellspace}
\usepackage{array}
\usepackage{dsfont}

\newcolumntype{C}{>{\centering\arraybackslash}X}
\newcolumntype{L}[1]{>{\raggedright\let\newline\\\arraybackslash\hspace{0pt}}m{#1}}

\newcommand{\indep}{\perp \!\!\! \perp}
\newcommand{\etal}{\textit{et al.}}

\begin{document}

\title{Comparing Causal Inference Methods for Point Exposures with Missing Confounders: A Simulation Study}


\author*[1]{\fnm{Luke} \sur{Benz}}\email{lukebenz@g.harvard.edu}
\author[2]{\fnm{Alexander W.} \sur{Levis}}
\author[1]{\fnm{Sebastien} \sur{Haneuse}}

\affil[1]{\orgdiv{Department of Biostatistics}, \orgname{Harvard T.H. Chan School of Public Health}, \orgaddress{\state{Massachusetts}, \country{U.S.A}}}

\affil[2]{\orgdiv{Department of Statistics \& Data Science}, \orgname{Carnegie Mellon University}, \orgaddress{\state{Pennsylvania}, \country{U.S.A}}}


\abstract{Causal inference methods based on electronic health record (EHR) databases must simultaneously handle confounding and missing data. In practice, when faced with partially missing confounders, analysts may proceed by first imputing missing data and subsequently using outcome regression or inverse-probability weighting (IPW) to address confounding. However, little is known about the theoretical performance of such reasonable, but \textit{ad hoc} methods. Though vast literature exists on each of these two challenges separately, relatively few works attempt to address missing data and confounding in a formal manner simultaneously. In a recent paper Levis \etal\cite{levis2022} outlined a robust framework for tackling these problems together under certain identifying conditions, and introduced a pair of estimators for the average treatment effect (ATE), one of which is non-parametric efficient. In this work we present a series of simulations, motivated by a published EHR based study\cite{arterburn2020} of the long-term effects of bariatric surgery on weight outcomes, to investigate these new estimators and compare them to existing \textit{ad hoc} methods. While methods based on \textit{ad hoc} combinations of imputation and confounding adjustment perform well in certain scenarios, no single estimator is uniformly best. We conclude with recommendations for good practice in the face of partially missing confounders.}




\maketitle

\section{Introduction}\label{sec:intro}
Electronic health record (EHR) databases contain observational data on large populations collected from patient interactions with a healthcare system over a potentially long time period. Given the sheer quantity of information captured, EHR databases are increasingly seen as a useful data source for research across a variety of clinical and public health settings \cite{iom2009, hudson2017, nih2009}. While randomized control trials (RCT) are often considered the gold standard, they may be infeasible to conduct in certain scenarios due to financial cost associated with running the trial, ethical considerations of randomizing subjects to a potentially harmful or inferior treatment, and/or other logistical concerns. Similar financial constraints and external factors may also make it infeasible to conduct prospective observational studies. In such scenarios when RCTs or prospective observational studies may not be possible, EHR data may allow for headway to be made towards studying the research questions of interest.

Despite their upsides, EHR are not without their own unique challenges that threaten the validity of any statistical analyses conducted using such data. Unlike data in a clinical trial, data in an EHR system are not collected with a given research purpose in mind, but rather are collected to record clinical activity and assist with patient billing. In particular, treatments which patients may receive are not randomly assigned, allowing for the presence of confounding bias\cite{hernan2020}. Similarly, information researchers would like to know about a patient may not be recorded at particular visits, or at all, meaning missing data is a fundamental challenge with which researchers must contend\cite{haneuse2016a}. Challenges in EHR-based observational studies rarely exist in isolation, and it is frequently the case that certain covariates necessary to address confounding may be unavailable for some patients. In this work, we consider the setting where interest lies in estimating the causal average treatment effect (ATE) of some point exposure on a given outcome in an EHR-based observational study where some of the confounders are missing for a subset of patients. 

While both confounding and missing data have their own extensive literatures, relatively few methods have been developed aiming to address both issues simultaneously within a formal framework, with notable exceptions including the recent work of Levis and colleagues\cite{levis2022}, and several works preceding it\cite{kennedy_2020, williamson2012, evans_2020, chen2007, seaman2014, sun2023identification, MSHB_shen_2023}. 

\begin{figure}[h]
\centering
   \includegraphics[scale=0.40]{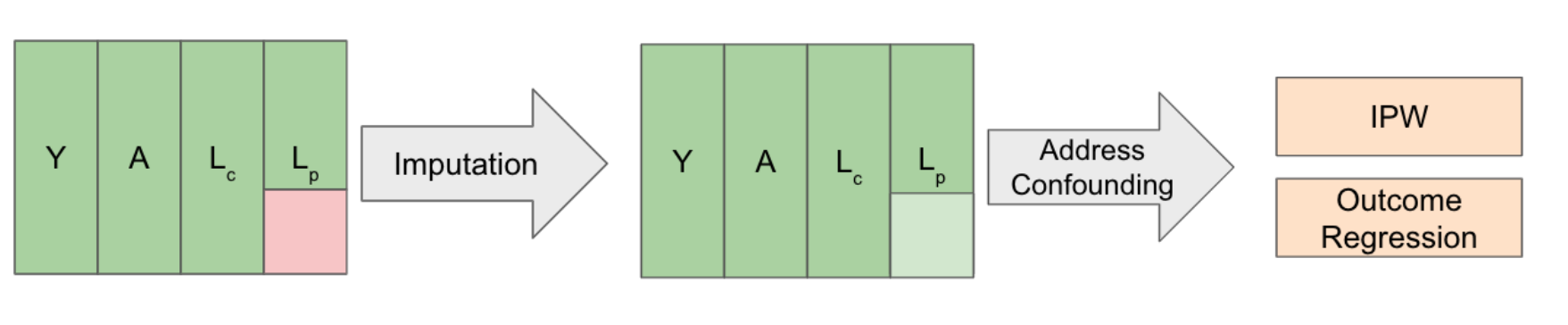}
\caption{Example analysis pipeline that decomposes the challenges of missing data and confounding as separate tasks. In this example, the analyst first applies imputation to address missing data and then applies their preferred method to deal with confounding.}
\label{fig:reasonable_analyst}
\end{figure}

In practice, when faced with the problem of partially missing confounders, analysts are likely to employ some \textit{ad hoc} combination of well known techniques for dealing with missing data and confounding separately, as outlined in Figure \ref{fig:reasonable_analyst}. For example, one might first use (multiple) imputation to deal with missing data and then feed the resulting imputed dataset(s) into a regression-based analysis to handle confounding. In certain cases such \textit{ad hoc} approaches may be perfectly reasonable and yield valid results, while in other cases such approaches will not yield valid results. 

Levis \etal~outlined a formal framework for estimating the causal average treatment effect from cohort studies when confounders satisfy a version of the missing at random (MAR) assumption, and proposed a pair of new estimators for this setting, one of which is an efficient and robust influence function-based estimator that serves as a theoretical benchmark against which such \textit{ad hoc} methods may be assessed. However no attempt was made to compare the performance of these new estimators to a number of reasonable strategies analysts might attempt in this setting. 

In this paper, we seek to use simulations to compare the performance of a number of reasonable approaches one may take when faced with partially missing confounders to the benchmarks outlined in Levis \etal\cite{levis2022}, so as to better understand under which scenarios such approaches may be valid and under which scenarios they may be invalid. The simulation settings considered in this work are motivated by the study of long-term outcomes following bariatric surgery. Despite the fact that several clinical trials and observational studies present evidence to suggest that bariatric surgery is the most effective weight-loss treatment for patients with extreme obesity\cite{o2013long, puzziferri2014long, gloy2013bariatric, sheng2017long, chang2014effectiveness}, the number of patients undergoing bariatric surgery each year is low\cite{arterburn2014bariatric}. Perhaps that relatively few patients are recommended to undergo bariatric surgery is due to the lack of evidence regarding long-term adverse events, durability of weight loss, and glycemic benefits, especially in comparison to patients who similarly suffer from severe obesity but do not undergo surgery\cite{o2013long}. In an ideal world, large scale randomized trials would be conducted to compare long-term outcomes between patients undergoing bariatric surgery to those not undergoing bariatric surgery. Such long-term studies, however, are currently unrealistic and would be slow to generate useful results. Therefore, rigorous and valid EHR-based studies may be the only feasible approach to generate evidence that can inform decisions regarding the long-term effects of bariatric surgery. As such, for these applications and beyond, it will be important to ascertain the operating characteristic of various methods intended to simultaneously handle missing data and confounding bias in these contexts.

\section{Background}\label{sec:background}
\subsection{Setting and Notation}\label{sec:setting}
The problem considered in this paper lies in using data from EHR databases to compare the effectiveness of a finite set of treatment options $A \in \mathcal{A}$ on some univariate outcome, which we denote $Y \in \mathcal{Y} \subseteq \mathbb{R}$. To be concrete, the motivating example we later base our simulation study on compares a set of bariatric surgery procedures ($\mathcal{A}$) on weight loss 5 years post surgery ($Y$). Let $Y(a)$ denote the counterfactual outcome corresponding to treatment $A = a$. In particular, we are interested in both estimation and inference of mean counterfactual outcomes $\mathbb{E}[Y(a)]$. In the setting where $|\mathcal{A}| = 2$ (i.e. the treatment is binary), we are interested in estimating the average treatment effect $\mathbb{E}[Y(1) - Y(0)]$. 

Given that such estimation and inference is to be conducted using observational EHR data, treatment is not randomly assigned and thus confounding must be addressed. Following the notation of Levis \etal\cite{levis2022}, we let $L \in \mathbb{R}^d$ denote a sufficient set of confounders that have been identified through a combination of expert knowledge and previous research efforts. In an ideal world, $L$ is fully observed for all subjects, and mean counterfactual outcomes can be estimated using standard causal methods that adjust for these confounders. In EHR, we are rarely so lucky, and instead must deal with $L = (L_c, L_p) \in \mathbb{R}^{s}\times \mathbb{R}^{q}$, where $L_c$ denotes confounders that are completely observed for all subjects, while $L_p$ denotes confounders that are only partially observed (missing for some subjects). For example, in our motivating bariatric surgery example, perhaps sex, race, and baseline BMI are available for every subject in the EHR database, but other important confounders such as comorbidities, and smoking status may only be ascertained for a subset of the subjects.

Let $R \equiv (R_1, \ldots, R_q) \in \{0, 1\}^q$ be a vector of binary indicators denoting whether or not each of the $q$ partially observed confounders $L_p$ is observed (i.e., $R_j = 1$ if and only if the $j$-th component of $L_p$ is observed), with $L_p^{(r)}$ representing the components of $L_p$ that are actually observed under the observation pattern $R = r$. As in Levis \etal\cite{levis2022}, we take the \textit{observed data} to be $n$ i.i.d. observations of $O = (L_c, A, Y, R, L_p^{(R)}) \sim P_o$. Defining $S = \mathds{1}(R = \bm 1_q)$ to be a complete data indicator, we can also consider the \textit{coarsened observed data}, $O' = (L_c, A, Y, S, SL_p) \sim P_{o'}$. When $S = 0$, we conceive of $SL_p = 0$ to denote that the entire $L_p$ is treated as missing (e.g., \texttt{NA})  Note that both $P_{o}$ and $P_{o'}$ are determined jointly by $P_f$, the distribution of the \textit{full data} $(L, A, Y)$, and the missingness mechanism that underpins $R$ \cite{levis2022}.

\subsection{Common Estimation Procedures}\label{sec:common_procedures}
When faced with the setting of partially missing confounders, there are several approaches an analyst might take if trying to estimate mean counterfactual outcomes. As shown in Figure \ref{fig:reasonable_analyst}, the challenge of dealing with missing data and confounding simultaneously often decomposes into two separate tasks for analysts: imputation of missing data, and then estimation of mean counterfactual outcomes on the imputed dataset(s), using methods such as outcome regression or inverse probability weighting to address confounding. 

Of course, analysts could in theory choose not to deal with the problem of missingness, and instead conduct a complete case analysis, which drops subjects who have missing data. Complete-case analyses are in general justified based on on the assumption that data are missing completely at random (MCAR), which is often unrealistic in observational EHR-based studies \cite{perkins2018}. Even if the MCAR assumption is met, a complete-case analysis may suffer efficiency loss due to excluding potentially valuable information on subjects with partially missing data \cite{perkins2018, allison2001, schafer1997}. Despite the obvious limitations of restricting analysis to only include complete cases, it is the most basic analysis technique and remains pervasive in epidemiological studies\cite{perkins2018, stuart2009, vanderheijden2006, westreich2012}. As such, we choose to include it as a comparison point in our simulation studies described in Section \ref{sec:simulations}.

\begin{center}
\begin{table}[ht]%
\centering
\begin{tabularx}{\textwidth}{ScccC}
\toprule
\textbf{Nuisance Function} & \textbf{Definition}  & \textbf{Description} & \textbf{Usage} \\
\midrule
$\lambda(\ell_p ~|~ L_c, A, Y, S = 1)$ & $p_{o'}[\ell_p~|~ L_c, A, Y, S = 1]$ & Imputation Model & Imputation of Missing Confounders \\
$\tilde\mu(y~|~ L_c, L_p, A)$ & $\mathbb{E}_{f}[Y  ~|~ L_c, A, L_p]$ & Outcome Model & Outcome Regression \\
$\tilde{\eta}(L_c, L_p, a)$ & $P_{f}[A = a ~|~ L_c, L_p]$ & Treatment Model & Propensity Score for Inverse Probability Weighting \\
\bottomrule
\end{tabularx}\caption{Summary of nuisance functions which may be estimated by analysts as they develop a strategy to estimate mean counterfactual outcomes under the combined setting of missing data and confounding. Analysts may first choose to impute missing data, and then apply outcome regression or IPW to estimate mean counterfactual outcomes in a manner that addresses confounding, although they could also apply these methods without imputing data (complete case analysis).\label{tab:common_causal_functions}}
\end{table}
\end{center}

Without \textit{a priori} knowledge of the missingness mechanism, imputation of missing data in our setting entails modeling $\lambda(\ell_p ~|~ L_c, A, Y, S = 1)$, the conditional distribution of missing confounders given the remaining observables, and using that distribution to fill in missing values by sampling from $\lambda$. Finally, whether or not analysts choose to impute missing values, they must contend with the issue of confounding. Reasonable choices of method might include outcome regression $\tilde \mu(y~|~ L_c, L_p, A) = \mathbb{E}_{f}[Y  ~|~ L_c, A, L_p]$ where $L_p$ may be imputed. Another alternative is inverse-probability weighting (IPW)\cite{sun2018, horwitz1952, shiba2021}, which weights subjects inversely proportional to the probability of receiving treatment $A = a$ conditional on all confounders $\tilde{\eta}(L_c, L_p, a) = P_{f}[A = a ~|~ L_c, L_p]$. Table \ref{tab:common_causal_functions} summarizes these nuisance functions which may be estimated by analysts as they develop a strategy to estimate mean counterfactual outcomes under the combined setting of missing data and confounding.

\subsection{Estimators Based on a Complete-Case MAR Assumption}\label{sec:levis_estimators}

In contrast to methods which address missing data and confounding separately, Levis and co-authors\cite{levis2022} tackle both issues jointly. Such a method is based on the following novel factorization of the coarsened observed data likelihood, where each braced quantity corresponds to a nuisance function that can be estimated from the observable data:

\begin{equation}\label{eqn:levis_factorization}
    p(O') = p(L_c)\underbrace{p(A~|~L_c)}_{\eta}\underbrace{p(Y~|~L_c, A)}_{\mu}\underbrace{p(S~|~L_c, A, Y)}_{\pi}\underbrace{p(L_p~|~L_c, A, Y, S = 1)^S}_{\lambda}.
\end{equation}

\begin{figure}[h]
     \centering
     \begin{subfigure}{0.49\textwidth}
         \centering
         \includegraphics[width=\textwidth]{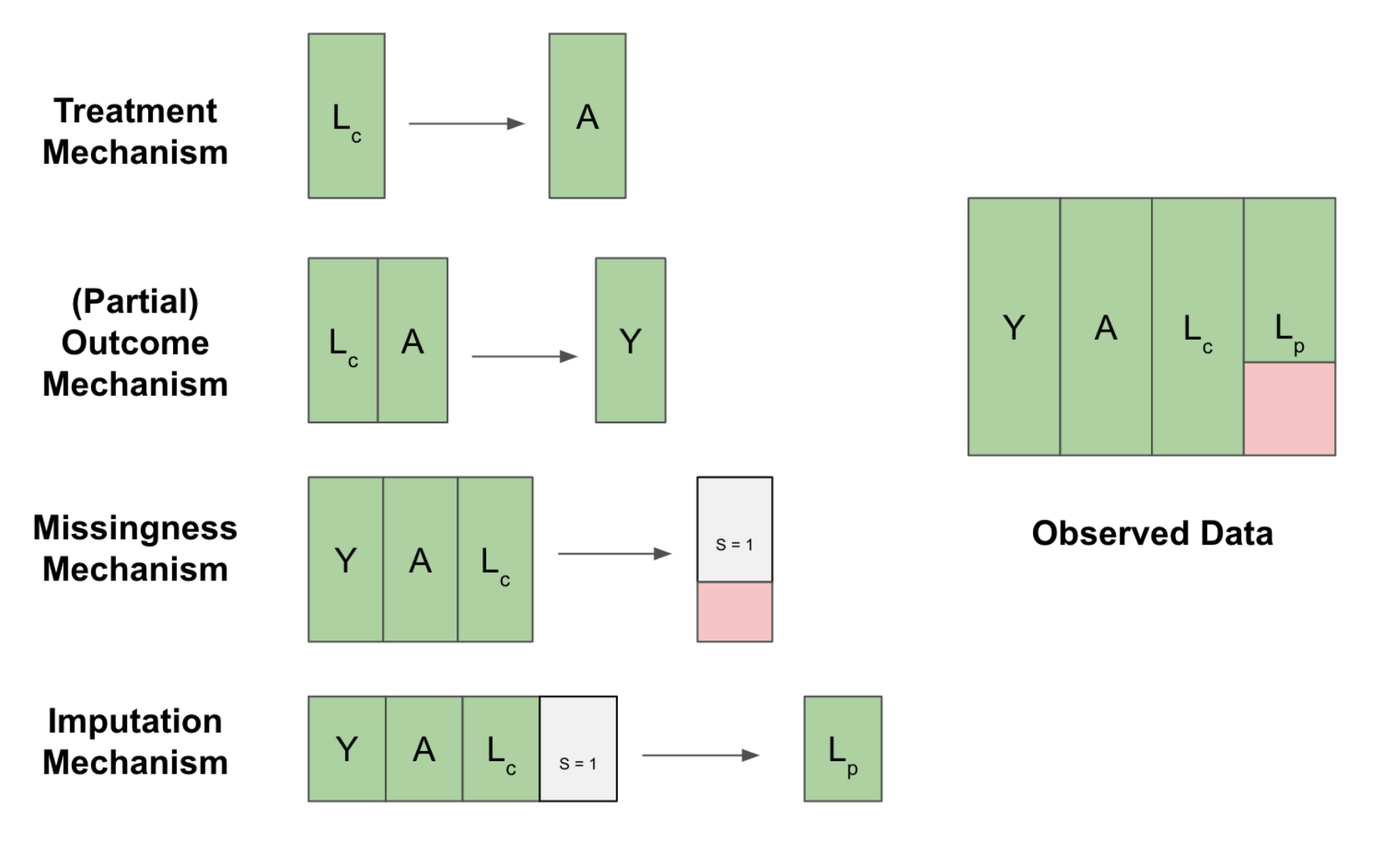}
         \caption{Levis Factorization}
         \label{fig:levis_factorization}
     \end{subfigure}
     \hfill
     \begin{subfigure}{0.49\textwidth}
         \centering
         \includegraphics[width=\textwidth]{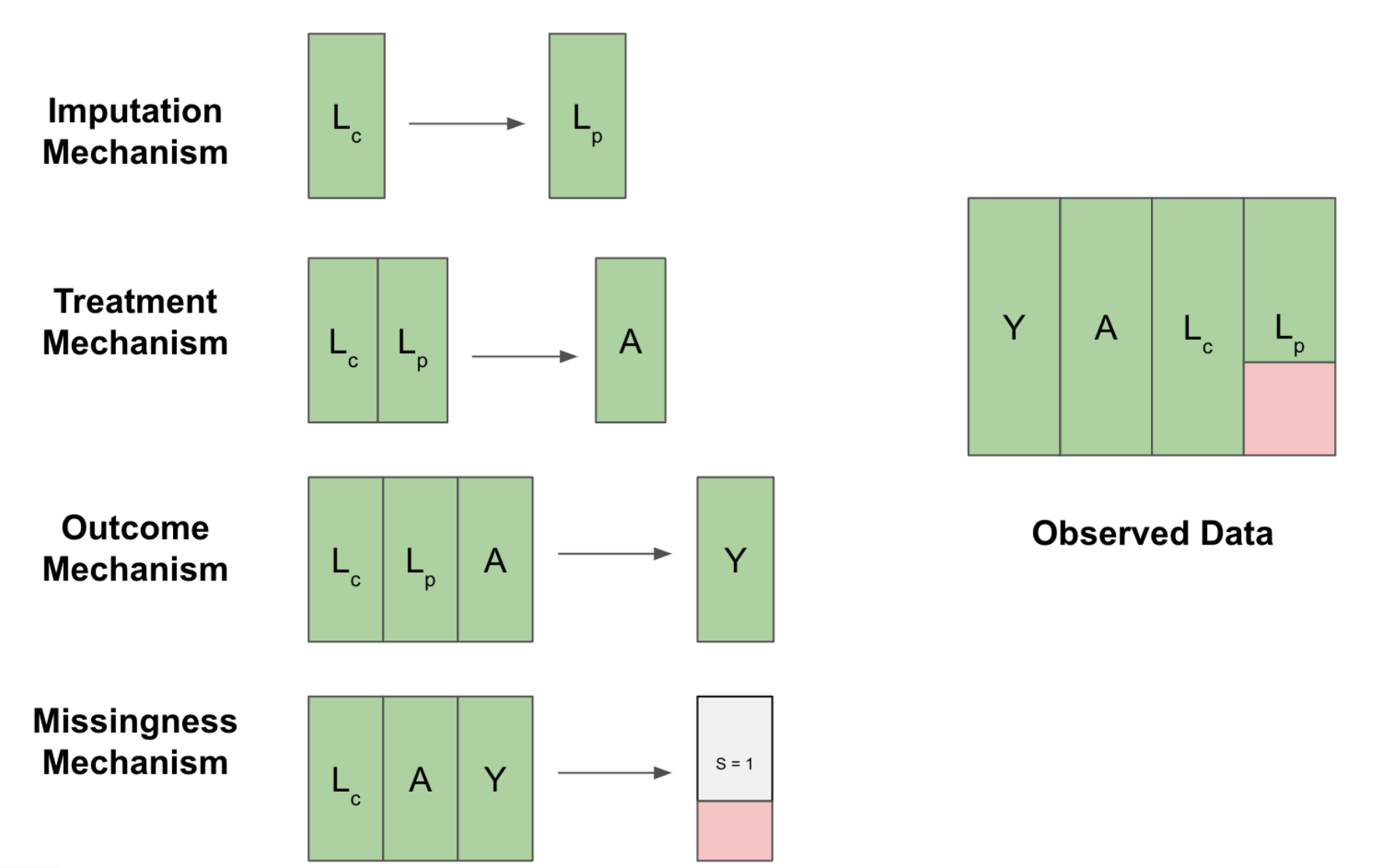}
         \caption{Alternative Factorization}
         \label{fig:alt_factorization}
     \end{subfigure}
        \caption{Two factorizations of the (coarsened) observed data likelihood. While the alternative factorization (shown in panel (b)) may seem more consistent with how data arise temporally, component nuisance models are not directly estimable from the observed data, whereas they are under the ``Levis factorization'' (shown in panel (a)) from Equation~\eqref{eqn:levis_factorization}.}
        \label{fig:factorizations}
\end{figure}

To paraphrase Levis \etal\cite{levis2022}: $\eta(L_c, a) = P_{o'}[A = a~|~ L_c]$ is a model for treatment mechanism conditional on only the subset of confounders that are observed for all subjects; $\mu(y~|~L_c, A) = P_{o'}[Y \leq y~|~L_c, A]$ is a outcome model based only on the treatment and the subset of confounders that are observed for all subjects; $\pi(L_c, A, Y) = P_{o'}[S = 1~|~L_c, A, Y]$ is a model for the probability of being a ``complete case'', which only depends on treatment, outcome, and complete confounders -- quantities that are observable for all subjects; and finally $\lambda(\ell_p~|~L_c, A, Y, S = 1)$ is the conditional density of $L_p$ given the remaining observed variables, among complete cases for which no confounders are missing. The $S$ in the exponent above the $\lambda$ term in Equation \eqref{eqn:levis_factorization} sets that portion of the coarsened observed data likelihood to $1$ if $L_p$ are not observed. Table \ref{tab:nuisance_functions} reproduces Table 1 from Levis \etal\cite{levis2022}, summarizing these nuisance functions. A visual overview of this novel data factorization is given in Figure \ref{fig:levis_factorization}.

On the basis of this novel factorization of the observed (coarsened) data likelihood, Levis \etal\cite{levis2022} introduce a complete-case missing at random (CCMAR) assumption, and propose two estimators for the mean counterfactual $\mathbb{E}[Y(a)]$ that are valid under CCMAR. The first is an inverse-weighted outcome regression (IWOR) estimator, $\tilde\chi_a$, and the second is an influence function (IF) based estimator $\hat\chi_a$. The influence function-based estimator, $\hat\chi_a$, is of primary interest as Levis \etal\cite{levis2022} are able to prove several results regarding the robustness and non-parametric efficiency of this estimator. Nevertheless, $\tilde\chi_a$ serves as a useful comparison point in our simulation study as another estimator derived from the factorization considered in Equation (\ref{eqn:levis_factorization}). Formal definitions and derivations of $\tilde\chi_a$ and $\hat\chi_a$ are presented in $\S5.1$ and $\S5.2$ of Levis \etal\cite{levis2022}, respectively; for completeness, we give the form of these estimators in the supplementary materials. Henceforth, we will refer to $\tilde{\chi}_a$ and $\hat{\chi}_a$ as CCMAR-based estimators. 

Formally, this complete-case missing at random assumption states that whether or not a patient is a ``complete case'' depends only on variables observed for all patients. That is, 
\begin{equation}\label{eqn:CCMAR}
S \indep  L_p \mid L_c, A, Y
\end{equation}
\noindent A noteworthy feature of this particular assumption is that it is agnostic to which components of $L_p$ may be missing for any particular subject. That is, this assumption is strictly in regards to whether a subject has complete data $(S = 1)$ or not $(S = 0)$. Of immediate consequence is that when this assumption holds, estimation strategies for $\tilde{\chi}_a$ and $\hat{\chi}_a$ do not require estimation or specification of complex, non-monotone missingness patterns that are common in EHR data\cite{haneuse2016a}. Possible limitations of this coarsening are discussed in detail in $\S7$ of Levis \etal\cite{levis2022}. A full set of identification assumptions for the ATE in this setting is covered in $\S3.1$ and $\S3.2$ of \cite{levis2022} and provided in our Supplementary Materials as well.

\begin{center}
\begin{table}[htbp]
\centering%
\begin{tabular}{ccccc}
\toprule
\textbf{Nuisance Function} & \textbf{Definition}  & \textbf{Description}  & $\tilde\chi_a$  & $\hat\chi_a$ \\
\midrule
$\eta(L_c, a)$ & $P_{o'}[A = a~|~ L_c]$ & Treatment Mechanism & & \checkmark \\
$\mu(y~|~L_c, A)$ & $P_{o'}[Y \leq y~|~L_c, A]$ & Outcome Distribution & \checkmark & \checkmark \\
$\pi(L_c, A, Y)$ & $P_{o'}[S = 1~|~ L_c, A, Y]$ & Missingness Mechanism &\checkmark & \checkmark \\
$\lambda(\ell_p ~|~ L_c, A, Y, S = 1)$ & $p_{o'}[\ell_p~|~ L_c, A, Y, S = 1]$ & Imputation Model & \checkmark & \checkmark \\
\bottomrule
\end{tabular}\caption{Summary of nuisance functions. Checkmarks indicate whether each component must be estimated for the influence function-based estimator $\hat\chi_a$ and the inverse-weighted outcome regression estimator $\tilde\chi_a$ from Levis \etal\cite{levis2022}\label{tab:nuisance_functions}}
\end{table}
\end{center}

The decomposition in Equation (\ref{eqn:levis_factorization}) may be considered somewhat unnatural given the temporal order of data occurring in the real world. Typically, when envisioning how data are generated, one first might imagine a process giving rise to any baseline covariates (confounders) of interest, followed by a mechanism for selection into a particular treatment group conditional on those covariates, and finally outcomes are generated conditional on covariates and treatment. Missing data might arise in a number of ways, but one may hypothesize an additional missing at random (MAR) missingness mechanism conditional on treatment, outcome, and observed covariates.

Such a factorization of the coarsened observed data likelihood, which we term ``alternative factorization'' may be expressed as follows.

\begin{equation}\label{eqn:alt_factorization}
    p(O') = p(L_c)\underbrace{p(L_p~|~L_c)^S}_{\tilde{\lambda}}\underbrace{p(A~|~L_c, L_p)}_{\tilde{\eta}}\underbrace{p(Y~|~L_c, L_p, A)}_{\tilde{\mu}}\underbrace{p(S~|~L_c, A, Y)}_{\pi}
\end{equation}

The components of this likelihood are slightly different than those in Equation \eqref{eqn:levis_factorization} in that what is being conditioned on is different. Nevertheless, we use similar notation to represent the nuisance functions in this alternative factorization. As shown in Figure \ref{fig:factorizations}, component nuisances functions under this factorization are not directly estimable from the observed data due to conditioning on partially observed confounders $L_p$.

\section{Data Driven Simulation Study}\label{sec:simulations}

In order to compare the validity of various reasonable but \textit{ad hoc} analytic approaches to the CCMAR-based estimators, we conduct a series of simulations. The aims of this simulation study are twofold:
\begin{enumerate}
    \item The primary goal of this simulation study is to learn where \textit{ad hoc} approaches that combine imputation with some method to account for confounding can perform reasonably well, and where they may break down. Because the CCMAR-based IF estimator, $\hat{\chi}_a$ has certain theoretical guarantees, benchmarking the performance of simpler approaches against the CCMAR-based estimators when data are generated under the factorization in Equation \eqref{eqn:levis_factorization} (e.g., when the nuisance models for CCMAR-based estimators are correctly specified) is of particular interest.
    \item A secondary goal of this simulation study is to evaluate the CCMAR-based estimators perform when data are generated by the alternative factorization in Equation \eqref{eqn:alt_factorization}. In such settings, the performance of CCMAR-based estimators is more closely aligned with how such estimators are likely to fare in real world settings where component nuisance models are unknown to the analyst.
\end{enumerate}

The framing of this simulation study is a hypothetical study comparing two bariatric surgery procedures on long term weight-loss maintenance \cite{arterburn2020}. Specifically, let $A$ be a binary point exposure taking on value 0 for Roux-en-Y gastric bypass (RYGB) and 1 for vertical sleeve gastrectomy (VSG). We let $Y$ denote the proportion weight change at 5 years post surgery, relative to baseline. In our simulations we consider up to 5 confounders: gender, baseline BMI at surgery (centered at 30), an indicator of Hispanic ethnicity, baseline Charlson-Elixhauser comorbidity score\cite{gagne2011}, and an indicator of smoking status; we denote these confounders as $( L^{(1)}, L^{(2)}, L^{(3)}, L^{(4)}, L^{(5)} )$, respectively.

To ensure our simulations reflect plausible real-world EHR-based settings, simulations are based on data from 5,693 patients who underwent either of the two bariatric procedures of interest at Kaiser Permanente Washington between January 1, 2008 and December 31, 2010. Complete information on gender, baseline BMI, and ethnicity, was available for all patients, while comorbidity scores were only available for 4,344 patients. Smoking status was not measured at all in the data, so we synthetically generated this confounder subject to partial missingness. This was done to create more complex scenarios with multiple partially missing confounders. In all simulation settings, the completely observed confounders were $L_c = ( L^{(1)}, L^{(2)}, L^{(3)})$, while the partially observed confounders were either the univariate $L_p=L^{(4)}$, or the bivariate $L_p = ( L^{(4)}, L^{(5)})$, depending on the simulation setting.

\subsection{Data Generation Process}\label{sec:DGP}
A critical component of our simulation infrastructure is the specification of the ``true'' components of the likelihood factorization corresponding to each respective aim. There is a subtle but important relationship between which factorization is used as the data generating process (DGP) and the series of nuisance functions required for each estimator. In particular, the nuisance functions required by the CCMAR-based estimators directly align with those under the factorization in Equation \eqref{eqn:levis_factorization} and the nuisance functions required by other ad-hoc estimators align with those under the factorization in Equation \eqref{eqn:alt_factorization}. When models required by each respective estimator align with components of the underlying DGP, it is straightforward to assess whether models are correctly specified because the ``true'' likelihood components are known. Similarly, when nuisance models required for an estimator do not directly correspond with likelihood components used to generate simulated data, it is not clear the degree to which component models may or may not be misspecified. 

The choice of which factorization to use as the DGP in each scenario was thus driven by which aim the scenario hoped to evaluate. Scenarios targeting aim 1 generated data from the novel factorization expressed in Equation \eqref{eqn:levis_factorization}, using CCMAR-based estimators using the ``true'' parametric models as a benchmark to compare a suite of approaches described in Section \ref{sec:methods}. Data generation under this factorization is described in Section \ref{sec:levis_factorization}. Scenarios targeting aim 2 compared estimators on data generated from the alternative factorization of the coarsened observed data likelihood in Equation \eqref{eqn:alt_factorization}, described in detail in Section \ref{sec:alternative_factorization} and shown visually in Figure \ref{fig:alt_factorization}. In such scenarios, the ``true'' nuisance functions in the factorization proposed by Levis \etal\cite{levis2022} (i.e., those in Table~\ref{tab:nuisance_functions}) were unknown and would not necessarily be derivable in closed form. As such, results of simulations under this specification can help demonstrate how the CCMAR-based estimators might perform when the true underlying nuisance functions are unknown. This, of course, is akin to the problem faced by an analyst in any practical setting. 

To inform specification of the ``true'' likelihood components for each factorization, a series of parametric models was fit to the Kaiser Permanente data as described in Sections \ref{sec:levis_factorization} and  \ref{sec:alternative_factorization}. This plasmode simulation set-up ensured that our set-up reflects the complexity of real EHR data to the extent possible by preserving complex correlation structures  \cite{Schreck2024Plasmode}.In each simulation setting, 5,000 datasets of size $n = 4,344$ patients were generated and analyzed according to the procedures outlined in the remainder of this section. In the main paper we present results from 4 data-driven simulations settings, summarized as follows.

\begin{enumerate}
    \item Factorization in Equation~\eqref{eqn:levis_factorization}, 1 partially missing confounder. 
    \item Factorization in Equation~\eqref{eqn:levis_factorization}, 1 partially missing confounder, additional non-linearities and interactions in imputation mean model,
    induced skew in distribution of imputation model.
    \item Factorization in Equation~\eqref{eqn:levis_factorization}, 2 partially missing confounders, amplified interactions between $A, L_{p}$, and $Y$ in joint imputation model, 
    induced skew in distribution of imputation model .
    \item Alternative factorization, 2 partially missing confounders, amplified interactions between $L_{c}, L_{p}$ in full treatment model $\tilde{\eta}(L_c, L_p, a)$.
\end{enumerate}
Data-driven simulation settings 1-3 target aim 1, while setting 4 targets aim 2. Results from an additional 15 data-driven simulation settings targeting both aims (and generating simulated data from both factorization) are available in the Supplementary Materials.

\subsubsection{Factorization in Equation~\eqref{eqn:levis_factorization}}\label{sec:levis_factorization}
To inform simulated datasets under Equation \eqref{eqn:levis_factorization}, each of the 4 nuisance functions for this factorization of the coarsened observed data likelihood, outlined in Table \ref{tab:nuisance_functions}, were modeled by fitting regressions to the Kaiser Permanente data. These models were used to help specify the ``true'' components of the likelihood. Specifically, logistic regressions were fit for $\eta$ and $\pi$, and a Gaussian linear model was fit for $\mu$. In the Kaiser Permanente data, the comorbidity score $(L^{(4)} = L_{p1})$ is categorical, but we instead chose to fit a generalized linear model with gamma link function to make the underlying relationship more complex. In fact, numerical integration techniques such as Gaussian quadrature are required to compute $\hat\chi_a$ when $L_p$ contains continuous confounders. 

In certain simulation settings, we simulated a second missing confounder, smoking status $(L^{(5)} = L_{p2})$ from a logistic regression model with pre-specified effect sizes given that no such smoker indicator is available in the Kaiser Permanente data. Such generation follows from the decomposition of the joint density of $L_p$ as follows

\begin{equation}\label{eqn:lp_joint}
\lambda(\ell_{p1}, \ell_{p2} ~|~ L_c, A, Y, S = 1) = \lambda_1(\ell_{p1} ~|~ L_c, A, Y, S = 1)\lambda_2(\ell_{p2} ~|~ L_c, A, Y, S = 1, L_{p1} = \ell_{p1})        
\end{equation}

\noindent where $\lambda_1$ and $\lambda_2$ represent the conditional densities of the 2 confounders in $L_p$, corresponding to $L^{(4)}$ and $L^{(5)}$, respectively.

In each simulation, $L_{c,1}...L_{c,n}$ were drawn with replacement from the empirical values in the Kaiser Permanente data. Then, $(A_i, Y_i, S_i, S_iL_{p,i})$ for $i = 1, ..., n$ were independently sampled, sequentially, from the following models, with covariates ($X$) and coefficients ($\beta$) for each model simulations available in Table \ref{tab:coeffients}.

\begin{itemize}
    \item $\text{logit}(\eta(L_c, 1)) = \text{logit}(P[A = 1~|~L_c]) = \beta_\eta^T X_\eta$
    \item $Y ~|~ L_c, A \sim \mathcal{N}(\beta_\mu^T X_\mu, \sigma^2_Y)$
    \item $\text{logit}(\pi(L_c, A, Y)) = \text{logit}(P[S = 1~|~L_c, A, Y]) = \beta_\pi^T X_\pi$
    \item $L_{p1}~|~L_c, A, Y, S = 1 \sim \text{Gamma}(\alpha, \lambda_1(L_c, A, Y))$ with rate parameter $\lambda_1(L_c, A, Y) = \frac{\alpha}{\mathbb{E}[L_{p1} ~|~ L_c, A, Y, S = 1]}$, and we parameterize $\log(\mathbb{E}[L_{p1} ~|~ L_c, A, Y, S = 1]) = \beta_{\lambda_1}^T X_{\lambda_{1}}$
    \item $\text{logit}(\lambda_2(L_c, L_{p1}, A, Y, S = 1)) = \text{logit}(P[L_{p2} = 1 ~|~ L_c, L_{p1}, A, Y, S = 1]) = \beta_{\lambda_2}^T X_{\lambda_2}$ (for simulations with multiple partially missing confounders).
\end{itemize}

We considered several settings where certain effects in the 4 nuisance models were amplified in order to create more interesting and complex relationships across confounders, treatment and outcome. 

\subsubsection{Alternative Factorization}\label{sec:alternative_factorization}
For aim 2, in order to asses the performance of CCMAR-based estimators where the ``true'' nuisance functions are unknown (and perhaps not derivable in closed form), we considered an additional data generating process based on Equation \eqref{eqn:alt_factorization} which reflects the more ``natural'' order described in the previous section. In simulations with multiple missing confounders, the joint density of $L_p$ decomposes as follows:

\begin{equation}\label{eqn:lp_joint_alt}
\tilde{\lambda}(\ell_{p1}, \ell_{p2} ~|~ L_c) =\tilde{\lambda}_1(\ell_{p1} ~|~ L_c, )\tilde{\lambda}_2(\ell_{p2} ~|~ L_c, L_{p1} = \ell_{p1})        
\end{equation}
\noindent where $\tilde{\lambda}_1$ and $\tilde{\lambda}_2$ represent the conditional densities of the 2 confounders in $L_p$, corresponding to $L^{(4)}$ and $L^{(5)}$, respectively.

Similar to the data-driven strategy described in Section~\ref{sec:levis_factorization}, 
in order to simulate datasets under Equation \eqref{eqn:alt_factorization}, each of the 4 nuisance functions for the alternative factorization of the coarsened observed data likelihood were fit to the Kaiser Permanente data to specify the ``true'' components of the likelihood. In fact, the type of parametric models used to fit these nuisance models under the alternative factorization were the same as under the previous factorization: logistic regressions were fit for $\tilde{\eta}$ and $\pi$, a Gaussian linear model was fit for $\tilde{\mu}$, a generalized linear model with gamma link function for $\tilde{\lambda}_1$ and a logistic regression for $\tilde{\lambda}_2$. The key differences between the models used to inform the truth in the previous factorization versus the alternative factorization are the sets of variables included in the relevant conditioning sets, i.e., what covariates can be conditioned on.

In each simulation, $L_{c,1}...L_{c,n}$ were drawn with replacement from the empirical values from the Kaiser Permanente data. Then $(L_{p,i}, A_i, Y_i, S_i)$ for $i = 1, ..., n$ were independently sampled, sequentially, from the following models.
\begin{itemize}
    \item $L_{p1}~|~L_c \sim \text{Gamma}(\alpha, \lambda_1(L_c))$ with rate parameter $\lambda_1(L_c) = \frac{\alpha}{\mathbb{E}[L_{p1} ~|~ L_c]}$, and we parameterize $\log(\mathbb{E}[L_{p1} ~|~ L_c]) = \beta_{\lambda_1}^T X_{\lambda_{1}}$
    \item $\text{logit}(\lambda_2(L_c, L_{p1})) = \text{logit}(P[L_{p2} = 1 ~|~ L_c, L_{p1}]) = \beta_{\lambda_2}^T X_{\lambda_2}$ (for simulations with multiple partially missing confounders).
    \item $\text{logit}(\eta(L_c, L_p, 1)) = \text{logit}(P[A = 1~|~L_c, L_p]) = \beta_\eta^T X_\eta$
    \item $Y ~|~ L_c, L_p, A \sim \mathcal{N}(\beta_\mu^T X_\mu, \sigma^2_Y)$
    \item $\text{logit}(\pi(L_c, A, Y)) = \text{logit}(P[S = 1~|~L_c, A, Y]) = \beta_\pi^T X_\pi$
\end{itemize}

Even though $S$ was generated after $L_p$, the model used to generate $S$ only conditions on $L_c, A, Y$, and thus the CCMAR assumption in Equation \eqref{eqn:CCMAR} is not violated. Because $L_p$ was sampled before $S$ in the above procedure, simulated values for $L_p$ were set to be missing among patients sampled to be non-complete cases ($S = 0$), thereby ensuring analysis took place on the same coarsened observed data structure $(L_c, A, Y, L_p, SL_p)$ as data generated under Equation \ref{eqn:levis_factorization}.

We considered several settings where certain effects in the 4 nuisance models were amplified in order to create more interesting and complex relationships across confounders, treatment and outcome. Covariates ($X$) and coefficients ($\beta$) for each model are presented in Table \ref{tab:coeffients}.

\begin{sidewaystable}
\footnotesize
\begin{minipage}[t]{0.5\textwidth}
\centering
\begin{tabular}[t]{>{}ccccc>{}c}
\hline
\multicolumn{1}{c}{\textbf{ }} & \multicolumn{1}{c}{\textbf{ }} & \multicolumn{4}{c}{\textbf{Data Driven Simulation Scenario}} \\
\cmidrule{3-6}
\textbf{Model} & \textbf{Term} & \textbf{1} & \textbf{2} & \textbf{3} & \textbf{4}\\
\hline
 & (Intercept) & -0.624 & -0.624 & -0.624 & -0.586\\
\cmidrule{2-6}
 & $L^{(1)}$ & 0.308 & 0.308 & 0.308 & 0.311\\
\cmidrule{2-6}
 & $L^{(2)}$ & -0.046 & -0.046 & -0.046 & -0.046\\
\cmidrule{2-6}
 & $L^{(3)}$ & -0.015 & 0.385 & 0.385 & 0.384\\
\cmidrule{2-6}
 & $L^{(4)}$ & --- & --- & --- & -0.035\\
\cmidrule{2-6}
 & $(L^{(2)})^2$ & 0.001 & 0.001 & 0.001 & 0.001\\
\cmidrule{2-6}
 & $L^{(1)} \times L^{(3)}$ & 0.400 & 0.400 & 0.400 & 0.100\\
\cmidrule{2-6}
\multirow{-8}{*}{\centering\arraybackslash $\eta$ or $\tilde{\eta}$} & $L^{(3)} \times L^{(4)}$ & --- & --- & --- & -0.020\\
\cmidrule{1-6}
 & (Intercept) & -0.207 & -0.207 & -0.207 & -0.200\\
\cmidrule{2-6}
 & $L^{(1)}$ & 0.031 & 0.031 & 0.031 & 0.032\\
\cmidrule{2-6}
 & $L^{(2)}$ & -0.002 & -0.002 & -0.002 & -0.002\\
\cmidrule{2-6}
 & $L^{(3)}$ & 0.023 & 0.023 & 0.023 & 0.022\\
\cmidrule{2-6}
 & $L^{(4)}$ & --- & --- & --- & -0.004\\
\cmidrule{2-6}
 & $L^{(5)}$ & --- & --- & --- & 0.200\\
\cmidrule{2-6}
 & $L^{(1)} \times L^{(3)}$ & -0.305 & -0.305 & -0.305 & -0.305\\
\cmidrule{2-6}
 & $L^{(4)} \times L^{(5)}$ & --- & --- & --- & -0.001\\
\cmidrule{2-6}
 & $A$ & 0.045 & 0.045 & 0.045 & 0.047\\
\cmidrule{2-6}
 & $A \times L^{(1)}$ & 0.413 & 0.313 & 0.313 & 0.313\\
\cmidrule{2-6}
 & $A \times L^{(2)}$ & -0.003 & -0.001 & -0.001 & 0.005\\
\cmidrule{2-6}
 & $A \times L^{(3)}$ & 0.130 & 0.080 & 0.080 & 0.081\\
\cmidrule{2-6}
 & $A \times L^{(4)}$ & --- & --- & --- & -0.002\\
\cmidrule{2-6}
 & $A \times L^{(5)}$ & --- & --- & --- & 0.100\\
\cmidrule{2-6}
\multirow{-15}{*}{\centering\arraybackslash $\mu$ or $\tilde{\mu}$} & $\sigma$ & 0.109 & 0.109 & 0.109 & 0.109\\
\cmidrule{1-6}
 & (Intercept) & 1.824 & 1.824 & 1.824 & 1.824\\
\cmidrule{2-6}
 & $L^{(1)}$ & 0.087 & 0.087 & 0.087 & 0.087\\
\cmidrule{2-6}
 & $L^{(2)}$ & 0.010 & 0.010 & 0.010 & 0.010\\
\cmidrule{2-6}
 & $L^{(3)}$ & -2.662 & -2.662 & -2.662 & -2.662\\
\cmidrule{2-6}
 & $L^{(2)} \times L^{(3)}$ & -0.149 & -0.149 & -0.149 & -0.149\\
\cmidrule{2-6}
 & $A$ & 2.922 & 2.922 & 2.922 & 2.922\\
\cmidrule{2-6}
 & $Y$ & 2.180 & 2.180 & 2.180 & 2.180\\
\cmidrule{2-6}
 & $A\times Y$ & 3.043 & 3.043 & 3.043 & 3.043\\
\cmidrule{2-6}
 & $A \times L^{(2)}$ & 0.159 & 0.159 & 0.159 & 0.159\\
\cmidrule{2-6}
\multirow{-10}{*}{\centering\arraybackslash $\pi$} & $Y \times L^{(1)}$ & 2.321 & 2.321 & 2.321 & 2.321\\
\hline
\end{tabular}
\end{minipage}
\begin{minipage}[t]{0.5\textwidth}

\begin{tabular}[t]{>{}ccccc>{}c}
\hline
\multicolumn{1}{c}{\textbf{ }} & \multicolumn{1}{c}{\textbf{ }} & \multicolumn{4}{c}{\textbf{Data Driven Simulation Scenario}} \\
\cmidrule{3-6}
\textbf{Model} & \textbf{Term} & \textbf{1} & \textbf{2} & \textbf{3} & \textbf{4}\\
\hline
 & (Intercept) & 0.867 & 0.939 & 0.939 & 0.900\\
\cmidrule{2-6}
 & $L^{(1)}$ & 0.075 & 0.074 & 0.074 & 0.272\\
\cmidrule{2-6}
 & $L^{(2)}$ & < 0.001 & -0.018 & -0.018 & 0.001\\
\cmidrule{2-6}
 & $L^{(3)}$ & -0.034 & -0.033 & -0.033 & -0.035\\
\cmidrule{2-6}
 & $(L^{(2)})^2$ & --- & < 0.001 & < 0.001 & ---\\
\cmidrule{2-6}
 & $L^{(1)} \times L^{(2)}$ & --- & --- & --- & 0.001\\
\cmidrule{2-6}
 & $L^{(1)} \times L^{(3)}$ & --- & --- & --- & 0.100\\
\cmidrule{2-6}
 & $A$ & 0.303 & 0.102 & 0.302 & ---\\
\cmidrule{2-6}
 & $Y$ & -0.765 & -0.696 & -0.696 & ---\\
\cmidrule{2-6}
 & $A\times Y$ & -0.500 & -0.050 & -0.500 & ---\\
\cmidrule{2-6}
 & $Y \times L^{(2)}$ & --- & -0.005 & -0.005 & ---\\
\cmidrule{2-6}
\multirow{-12}{*}{\centering\arraybackslash $\lambda_1$ or $\tilde{\lambda}_1$} & $\alpha$ & 3.619 & 1.000 & 1.000 & 3.616\\
\cmidrule{1-6}
 & (Intercept) & --- & --- & -1.386 & -1.386\\
\cmidrule{2-6}
 & $L^{(1)}$ & --- & --- & --- & 0.030\\
\cmidrule{2-6}
 & $L^{(2)}$ & --- & --- & --- & 0.050\\
\cmidrule{2-6}
 & $L^{(3)}$ & --- & --- & --- & 0.025\\
\cmidrule{2-6}
 & $L^{(4)}$ & --- & --- & 0.075 & 0.075\\
\cmidrule{2-6}
 & $A$ & --- & --- & --- & ---\\
\cmidrule{2-6}
 & $Y$ & --- & --- & -4.000 & ---\\
\cmidrule{2-6}
 & $A\times Y$ & --- & --- & 2.000 & ---\\
\cmidrule{2-6}
 & $A \times L^{(4)}$ & --- & --- & -0.010 & ---\\
\cmidrule{2-6}
 & $Y \times L^{(2)}$ & --- & --- & --- & ---\\
\cmidrule{2-6}
\multirow{-11}{*}{\centering\arraybackslash $\lambda_2$ or $\tilde{\lambda}_2$} & $Y \times L^{(4)}$ & --- & --- & 0.010 & ---\\
\hline
\end{tabular}
\end{minipage}
\caption{Coefficients ($\beta$) and covariates ($X$) used to generated simulated datasets for each of the four simulation scenarios presented in the main paper. Scenarios 1-3 generated datasets using the Levis factorization (Equation (\ref{eqn:levis_factorization})), while scenario 4 generated datasets using an alternative factorization (Equation (\ref{eqn:alt_factorization})). --- denotes the given covariate was not including in the corresponding model. Note that simulation scenarios 1-2 only generated a single partially missing confounder and as such no $\lambda_2$ model was needed.}
\label{tab:coeffients}
\end{sidewaystable}

\subsubsection{Summary of Scenarios}

Because of differences in the relative ordering of component models between the two factorizations, coefficients in Table \ref{tab:coeffients} are not directly comparable across scenarios using different factorizations. As such, we provide some additional summary information describing the four simulation scenarios. 

The true average treatment effects, reflecting the relative (\%) weight loss advantage of RYGB over VSG were 11.9\%, 10.4\%, 10.2\% and 22.7\%, respectively. Additional details on how these ground truth ATEs were computed is summarized in the Supplementary Materials (Section S4). The marginal probability of missingness, $p(S = 0)$, remained fairly consistent across scenarios 1-4, totaling 32.5\%, 30.8\%, 30.8\%, and 29.5\%, respectively. 

Unfortunately, we are unable to provide a single summary measure quantifying the magnitude of confounding directly attributable to $L_p$, as that is a complex function of the distribution of observed $L_p$, the strength of associations between $L_p, A$ and $Y$, and the functional form of $L_p$ in component nuisance models. Nevertheless, based on the dimension of $L_p$ and magnitude of coefficients in Table \ref{tab:coeffients}, it's reasonable to characterize the four simulation scenarios as increasing in terms of the complexity required to account for confounding due to $L_p$ (scenario 1 easiest $\to$ scenario 4 hardest).

\subsection{Methods}\label{sec:methods}

Across our simulation settings, we considered combinations of several methods for imputation of $L_p$ in conjunction with the use of outcome regression or inverse probability weighting to address the issue of confounding. Specifically, we considered methods that vary in both model flexibility (parametric, semi-parametric, non-parametric) and structure (interactions vs. no interactions). Taken collectivity, the range of methods considered reflects the types of models analysts might used in practice \cite{hernan2020}, and as such serves as a useful set to evaluate in aim 1.

Under each imputation strategy, a parametric model was estimated for $L_p$, and subsequently a single draw was sampled from the induced conditional distribution for each missing observation. The choice to use single imputation rather than multiple imputation helped ease computational given the large number of simulation scenarios and comparator methods considered. Three imputation models for $L_p$ were considered: the correctly specified DGP for $L_p$ (a gamma generalized linear model for comorbidity score $(L^{(4)})$ and a logistic regression for smoking status $(L^{(5)})$, with exact models depending on the corresponding factorization under which data were generated); a simpler imputation model with a Gaussian linear model for comorbidity score and a logistic regression for smoking status, with neither model considering any interactions; and, an imputation model with a Gaussian linear model for comorbidity score and a logistic regression for smoking status, while specifying all pairwise interactions. To avoid overfitting/variance inflation due to specifying all pairwise interactions, linear/logistic regression models in this third case were fit using \texttt{glmnet}\cite{glmnet} in \texttt{R} with $L_1$ (LASSO)\cite{lasso} regularization. Additionally, we considered a fourth strategy where the problem of missing data was ignored and subjects with missing these data were dropped, in order to mimic a complete case analysis.

Upon imputing (or not imputing) $L_p$, we considered several methods to directly model $Y$ in order to estimate the mean counterfactual outcome $\mathbb{E}[Y^{(a)}]$ using outcome regression. In particular, we considered 3 types of models ranging in flexibility: linear regression, with no interactions and with all pairwise interactions specified; generalized additive models (GAMs)\cite{gam2017}, with no interactions and with all pairwise interactions specified; and random forest. As was the case with the linear regression models used for imputation of $L_p$, outcome regression models specifying all pairwise interactions were estimated with $\texttt{glmnet}$\cite{glmnet} using $L_1$ (LASSO)\cite{lasso} regularization. 

In the case of GAMs, interactions took on one of three forms. When two covariates were binary, interaction effects were simply the coefficient for product of the covariates, as in a linear regression model. For covariates $x_1 \in \{0, 1\}, x_2 \in \mathbb{R}$, interaction effects were the coefficients for the product $x_1f(x_2)$, where $f$ is some smooth basis function. Finally, when both covariates were continuous, interaction effects were coefficients for the two dimensional surface $g(x_1, x_2)$. In our fitting of GAMs, $f$ was chosen to be a cubic regression spline, while $g$ was a tensor product smoother. Inherent to the fitting of these smooth functions is regularization in the form of choice of degree and/or knots\cite{wood2016}. Finally, we considered an even more flexible non-parametric random forest model, which can capture complex, highly non-linear relationships between confounders, treatment and outcome. 

We employed a similar suite of approaches for IPW as we did for outcome regression, instead modeling the probability of treatment using several models varying in complexity and flexibility. The analogue of the linear regression (with and without interaction effects) when modeling propensity scores was logistic regression, while GAMs with logit link (i.e., GAM logistic regression) were used as the analogue of GAM outcome regression. Random forests were also applied under classification specification to most flexibly model the treatment mechanism. Table \ref{tab:strawman_models} summarizes the suite of ``standard'' models.

Consistent with aim 1,  CCMAR-based estimators were computed using correctly specified parametric models in scenarios where data was generated under Equation~\eqref{eqn:levis_factorization} (scenarios 1--3). Towards evaluation of aim 2, we fit three versions of the CCMAR-based estimators in scenarios where data was generated under Equation \eqref{eqn:alt_factorization} (scenario 4). First, we fit the same parametric models used in scenarios 1-3, which we anticipate would be misspecificed. Given that the induced form of the correct nuisance functions under Equation \eqref{eqn:levis_factorization} was not easily derived in closed form when data were generated under Equation \eqref{eqn:alt_factorization}, we considered a flexible version of the CCMAR-based estimators leveraging GAMs with logit link for $\hat \eta$, $\hat \pi$ and $\hat \lambda_2$, a traditional GAM for $\hat \mu$, and a GAM with underlying gamma distribution for $\hat \lambda_1$. In this flexible version of the CCMAR-based estimators, GAMs include all pairwise interactions. We also considered a flexible version of the CCMAR-based estimators with a GAM with underlying Gaussian distribution for $\hat \lambda_1$.

\begin{table}[ht]
    \footnotesize
    \centering
    \begin{tabularx}{\textwidth}{Sc@{}Sc@{}Sc@{}C@{}C}
    \hline
    \multicolumn{5}{c}{\multirow{2}*{\textbf{Imputation of }$L_p$}} \\ \\
    \hline
     \textbf{Imputation Process} & \textbf{Confounder} & \textbf{Model Type} &  \textbf{Covariates} & 
     \textbf{Interactions} \\
     \hline
     \multirow{2}*{True Data Generating Process} & $L^{(4)}$ & Gamma GLM & $A, Y, L_c$ & \text{When Applicable} \\
     & $L^{(5)}$ & Logistic Regression & $A, Y, L_c, L^{(4)}$ & \text{When Applicable} \\ 
    \hline
    \multirow{2}*{Simple Imputation} & $L^{(4)}$ & Linear Regression & $A, Y, L_c$ & ---\\
    & $L^{(5)}$ & Logistic Regression & $A, Y, L_c, L^{(4)}$ & --- \\
    \hline
    \multirow{2}*{Imputation w/ Interactions} & $L^{(4)}$ & Linear Regression & $A, Y, L_c$ & \text{All Pairwise}$^*$\\
    & $L^{(5)}$ & Logistic Regression & $A, Y, L_c, L^{(4)}$ & \text{All Pairwise}$^*$ \\
    \hline
    \multirow{2}*{No Imputation (Complete Case)} & $L^{(4)}$ & ---& --- & --- \\
    & $L^{(5)}$ & --- & --- & ---\\

    \hline
    \multicolumn{5}{c}{\multirow{2}*{\textbf{Outcome Regression Model}}} \\ \\
    \hline
     \multicolumn{2}{c}{\textbf{Model Type}} & {\textbf{Covariates}} & \textbf{Imputed Datasets Used} & 
     \textbf{Interactions} \\
     \hline
     \multicolumn{2}{c}{\multirow{2}*{Linear Regression}} &  \multirow{2}*{$A, L_c, L_p$} & \multirow{2}*{All} & --- \\
      & & & & \text{All Pairwise}$^*$ \\ 
      \hline
     \multicolumn{2}{c}{\multirow{2}*{Generalized Additive Models}} &  \multirow{2}*{$A, L_c, L_p$} & \multirow{2}*{All} &  --- \\
      & & & & \text{All Pairwise} \\ 
      \hline
      \multicolumn{2}{c}{Random Forest} & $A, L_c, L_p$ & All &  Implicit in model \\

    \hline
    \multicolumn{5}{c}{\multirow{2}*{\textbf{Probability of Treatment Model (IPW) }}} \\ \\
    \hline
     \multicolumn{2}{c}{\textbf{Model Type}} & {\textbf{Covariates}} & \textbf{Imputed Datasets Used} & 
     \textbf{Interactions} \\
     \hline
     \multicolumn{2}{c}{\multirow{2}*{Logistic Regression}} &  \multirow{2}*{$L_c, L_p$} & \multirow{2}*{All} & --- \\
      & & & & \text{All Pairwise}$^*$ \\ 
      \hline
     \multicolumn{2}{c}{\multirow{2}*{\shortstack{Generalized Additive Models\\ (Logistic Regression)}}} &  \multirow{2}*{$L_c, L_p$} & \multirow{2}*{All} &  --- \\
      & & & & \text{All Pairwise} \\ 
      \hline
      \multicolumn{2}{c}{Random Forest} & $L_c, L_p$ & All &  Implicit in model \\
    \hline
    \end{tabularx}
    \begin{tablenotes}
      \footnotesize
      \item[*] LASSO\cite{lasso, glmnet} regularization in model with pairwise interactions specified in order to select meaningful interaction effects.
    \end{tablenotes}
    \caption{Summary of approaches in simulation settings combining imputation with one of two methods for accounting for confounding (outcome regression or IPW). First, partially missing confounders $L_p$ are imputed using one of 3 techniques, or alternatively subjects missing these data are dropped to mimic a complete case analysis. Various types of models ranging in flexibility---both in terms of types of interactions included as well as the structure of the model---are estimated to model either the outcome or probability of treatment in order to estimate mean counterfactual outcomes $\mathbb{E}[Y^{(a)}]$ in a manner that addresses confounding.}
    \label{tab:strawman_models}
\end{table}

\section{Simulation Study with Non-Parametric Nuisance Models}\label{sec:np_sims}
The suite of simulation scenarios based on our running bariatric surgery example in Section \ref{sec:simulations} does not consider the performance of the CCMAR-based estimators of Levis \etal\cite{levis2022} using fully non-parametric model choices for the nuisance functions. In data driven simulation scenario 4, GAMs were used to flexibly model the mean parameter in each nuisance model, but GAMs still make distributional assumptions for the residuals (e.g. Gamma or Gaussian).  Given that nonparametric modeling choices might be appealing when little is known about the underlying data generating process, understanding the performance of CCMAR-based estimators in this setting is an important component of aim 2 and helps inform recommendation about the use of these estimators in practice.

In order to evaluate the performance of these new estimators in a fully flexible manner, we designed an additional, small scope simulation study based on the worked data example described in Levis \etal\cite{levis2022}, which utilized a fully non-parametric approach. Due to computational overhead associated with non-parametric conditional density estimation, and subsequent numerical integration over such densities, this simulation study is somewhat simpler than those presented in the previous section. Nevertheless, the presence of multiple missing confounders make it sufficiently challenging to gain insight into the performance of  CCMAR-based estimators (particularly the CCMAR-IF estimator) in their most flexible form.

\subsection{Data Generation Process}
In this simulation $L_c = \emptyset$ and $L_p = (L_{p1}, L_{p2})$, where $L_{p1}$ is a binary covariate and $L_{p2}$ is a continuous covariate. For 1,000 replicates of size $n = 5,000$ subjects, i.i.d. copies of $(A_i, Y_i, S_i, SL_{p,i})$ were generated as follows:

$$
\begin{aligned}
    A & \sim \text{Bernoulli}(0.5) \\
    Y~|~ A &\sim \begin{cases}\text{Beta}(2,4) & \text{if}~A = 0 \\ \text{Beta}(4,2) & \text{if}~A = 1\end{cases} \\
    \text{logit}(P[S = 1~|~A, Y]) &= -0.35 + 0.5A + 0.18 Y + 0.05A\times Y \\
    \text{logit}(P[L_{p1} = 1~|~ A, Y, S = 1]) &= -0.6 + 0.5A + 0.25 Y + 0.1A\times Y \\
    L_{p2} ~|~ A, Y, S = 1, L_{p1} &\sim \mathcal{N}(A + Y + 2.5L_{p1}\times Y, 1.25^2)
\end{aligned}
$$

\subsection{Methods}
The CCMAR-based estimators were computed using non-parametric methods for each nuisance function, as well as, for comparison, the correctly specified parametric nuisance functions. In the case of non-parametric estimation, $\hat\eta(0) = \hat\eta(1) = 0.5$, essentially assuming treatment assignment probabilities were known by design. The conditional density of $Y$ given $A$, $d\hat{\mu}$ was fit using the highly adaptive lasso conditional density estimator\cite{hejazi2022efficient} in the \texttt{haldensify} \texttt{R} package \cite{hejazi2022haldensify-joss, hejazi2022haldensify-rpkg}, with 25 equally sized bins and maximum degree of basis function interaction of 3. Probability of missingness $\hat \pi(A, Y)$ was estimated with an ensemble regression learner from the \texttt{SuperLearner} package in \texttt{R}\cite{van2007super, superlearner}. Component models in the ensemble included a generalized linear model with all second order interactions, as well as multivariate adaptive polynomial spline regressions\cite{polspline}. The probability mass function of the binary covariate $L_{p1}$, $\hat{\lambda}_1(\ell_{p1}~|~A, Y)$, was estimated via the same \texttt{SuperLearner} ensembling strategy. Finally, the conditional density of the continuous partially observed covariate $\hat\lambda_2(\ell_{p2}~|~A, Y, \ell_{p1})$ was estimated using the same highly adaptive lasso conditional density estimator that was used for estimating $d\hat\mu$. 

Due to the non-parametric nature of the nuisance function models, the CCMAR-based estimators were computed via 2-fold cross fitting\cite{zheng2011, bickel1982, chernozhukov2018}. Briefly, this involved splitting each simulated data set into two equally sized partitions, using one to fit the nuisance models and the other to compute the estimator on the basis of the estimated nuisance functions. The roles of the two folds were then swapped and the results from each fold were averaged to produce a final estimator per dataset.

\section{Results}\label{sec:results}
Results from the 4 most pertinent simulation scenarios based on the bariatric surgery example are presented in Tables \ref{table:results_1}--\ref{table:results_4}. Results examining the properties of the fully non-parametric CCMAR-based estimators are presented in Table \ref{table:results_ff}. Expanded results from an additional 15 data driven simulation scenarios are available in our Supplementary Materials. We summarize several key findings and observations as follows:

\begin{figure}
\centering
   \includegraphics[width = \textwidth]{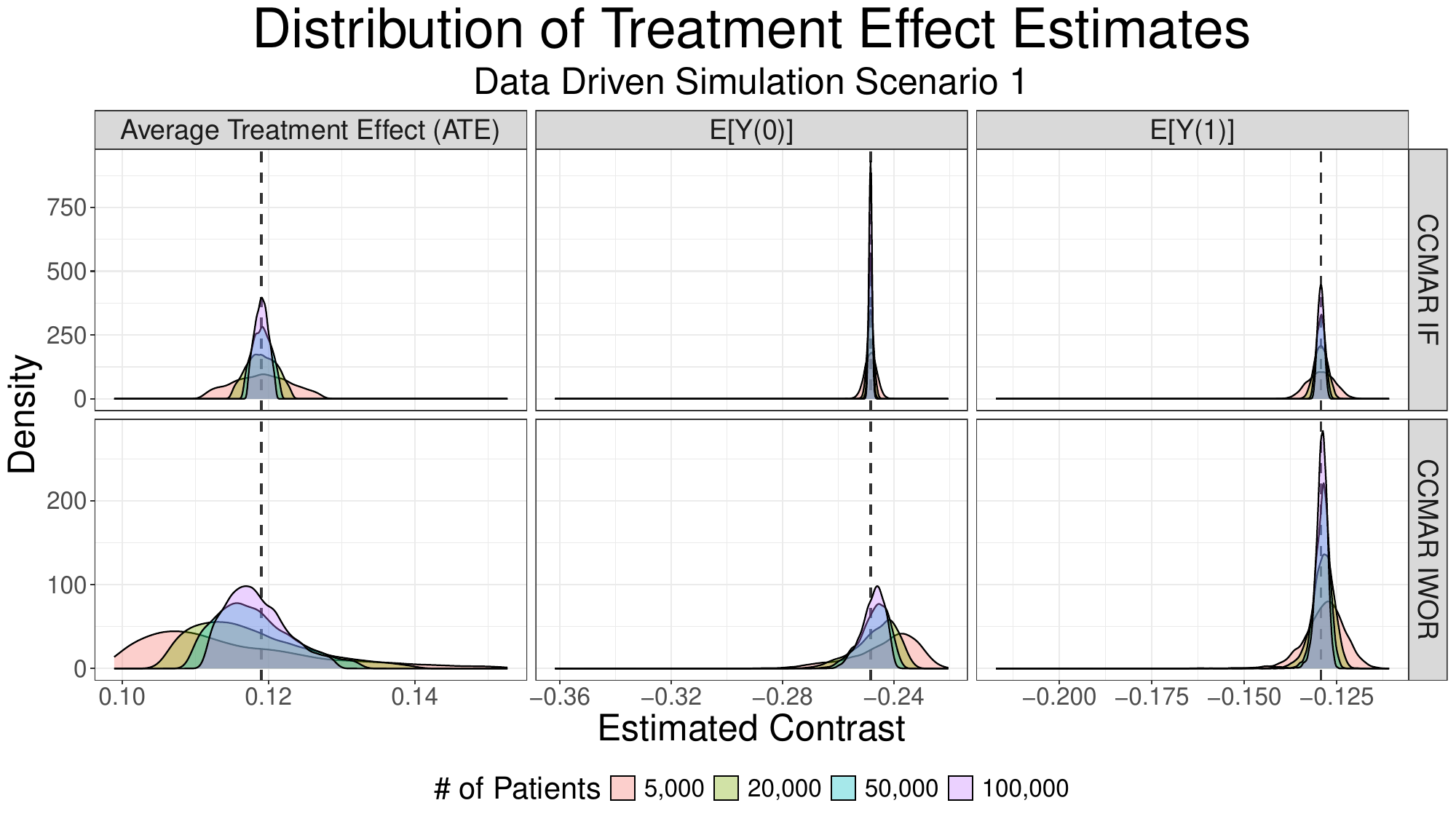}
\caption{Distribution of estimated treatment effects under simulation scenario 1 using the complete-case missing at random (CCMAR) influence function-based (IF) and inverse-weighted outcome regression (IWOR) estimators. 5,000 simulated datasets where generated for each number of patients, $n$. For every contrast, the IWOR estimator does tend toward asymptotically normal, but its empirical rate of convergence appears to be much slower than the IF based estimator. This behavior is likely what gives rise to the larger median bias observed in the IWOR estimator, despite the mean having little to no bias in most scenarios.}
\label{fig:levis_skew}
\end{figure}

\begin{enumerate}
    \item\textbf{Complete case analysis can be severely biased}: In every scenario considered, dropping subjects with missing values for some of the confounders lead to relative bias on the order of magnitude of 20\%. This held regardless of whether outcome regression or inverse probability weighting methods were employed, whether flexible models were used, and whether interactions were included in the models. 
\end{enumerate}

\begin{enumerate}
    \setcounter{enumi}{1}
    \item \textbf{Sufficient model flexibility can overcome confounding bias due to model misspecification}: As seen in scenario 1, adding pairwise interactions to outcome regression models (OLS and GAM) or propensity score models (logistic regression, GAM logistic regression) removed bias observed when such terms were omitted. Similarly, random forest models for either the outcome or propensity score yielded results with very small bias.
\end{enumerate}

\begin{enumerate}
    \setcounter{enumi}{2}
    \item \textbf{Flexibility does not always come at the expense of efficiency}: In the majority of scenarios considered, the relative uncertainty of methods not including interactions exceeded 1, indicating estimators were less efficient than the influence function CCMAR-based estimator. In fact, including pairwise interaction terms frequently improved estimator efficiency. In many scenarios, such as \# 1-2, the inclusion of pairwise interactions pushed relative uncertainty below 1, indicating estimators were more efficient than the influence function CCMAR-based estimator. Additionally, non-parametric random forest methods consistently displayed some of the best efficiency metrics observed across scenarios. Thus, perhaps somewhat counter to common intuition, efficiency is not always lost when increasing model flexibility.
\end{enumerate}

\begin{enumerate}
    \setcounter{enumi}{3}
    \item \textbf{Model flexibility without care does not guarantee unbiasedness}: Deploying flexible models, like GAMs, without interactions yielded bias in both directions depending on the simulation scenario and analysis method (IPW versus outcome regression). In fact, without the inclusion of interactions, flexible models like GAM did no better than simple OLS. This may be because with cubic regression splines, GAMs may regularize many terms to first order anyway. Additionally, GAMs can account for non-linearities but does not inherently account for interactions unless such terms are specified. Similarly, even fully non-parametric models such as random forests, which can account for both non-linearities and complex interactions in predictors, yielded non-trivial bias in scenarios 2-4. Thus, while we encourage consideration of flexible modeling choices, we caution default settings and acknowledge that such models typically require additional attention to (hyper)parameters compared to simpler modeling choices.
\end{enumerate}

\begin{enumerate}
    \setcounter{enumi}{4}
    \item \textbf{Imputation method less important when Normal distribution approximates} $\lambda_{1}$ \textbf{well}: In scenario 1, where the single partially missing confounder followed a gamma distribution, there is hardly any difference in bias/efficiency of standard methods across imputation methods. That is, Gaussian linear regression, with or without interactions, seems to impute $L^{(4)}$ sufficiently well such that analysis of corresponding datasets yields results very similar to analysis of datasets where $L^{(4)}$ was imputed using the true gamma data generating process. In scenario 1, the Gamma shape parameter of $\alpha = 3.619$ implies that the distribution of $L^{(4)}$ is not too skewed and can be decently approximated by a Normal distribution. However, in scenarios 2-3 where $\alpha = 1$ was selected to induce additional skew in the distribution of $L^{(4)}$, unbiased results only arise when the imputation process is in line with the true data generating process, based on the correctly specified Gamma distribution.
\end{enumerate}

\begin{enumerate}
    \setcounter{enumi}{5}
    \item \textbf{Standard methods perform well even with multiple missing confounders and amplified relationships}: Many of the same patterns emerge in more complex scenarios (e.g., \#3) where there are multiple partially missing confounders as well as amplified interaction effects between $A, Y$ and $L_p$. Namely: 
    \begin{itemize}
        \item Methods using linear/logistic regressions and GAMs have some bias when interaction terms are not specified.
        \item Bias is attenuated when specifying interaction terms in methods using linear/logistic regressions and GAMs.
        \item Efficiency tends to improve when including interaction terms.
        \item Most methods have relative uncertainty $< 1$ (better than influence function CCMAR-based estimator).
        \item There exist scenarios where every method is slightly biased unless true data generating process is used for imputation (scenario \#3).
    \end{itemize}
\end{enumerate}

\begin{enumerate}
    \setcounter{enumi}{6}
    \item \textbf{Influence function CCMAR-based estimator can be biased when nuisance functions misspecified}: Under the alternative data generating processes (scenario \#4) the CCMAR-based estimators are biased when (incorrect) parametric models are used to estimate nuisance functions. In scenario 4, the majority of standard methods also are biased to varying degrees, as their nuisance model specifications may also be incorrect, but the influence function CCMAR-based estimator has relative bias greater than any standard method outside of complete case analyses.
\end{enumerate}

\begin{enumerate}
    \setcounter{enumi}{7}
    \item \textbf{Influence function CCMAR-based estimation with flexible modeling of nuisance functions can overcome bias due to misspecification}: In scenario 4, the flexible version of the influence function CCMAR-based estimator, with semi-parametric (GAM) specification of nuisance functions yielded nearly unbiased estimation of the average treatment effect. Flexibly modeling the nuisance functions in Equation \eqref{eqn:levis_factorization} fully addressed the bias observed from the parametric specification of the influence function CCMAR-based estimator. Furthermore, as seen in Table \ref{table:results_ff}, this estimator with fully non-parametric nuisance models achieved unbiased results, demonstrating the success of this estimator even in the absence of any knowledge about the true data generating process.
\end{enumerate}

\begin{enumerate}
    \setcounter{enumi}{8}
    \item \textbf{The IWOR CCMAR-based estimator is skewed and has poor relative uncertainty}: In most scenarios, the CCMAR-based inverse-weighted outcome regression (IWOR) estimator has non-trivial relative bias in the median, even if the mean is unbiased. This skewness can be seen in Figure \ref{fig:levis_skew}, and compared to the influence function-based estimator, convergence to asymptotic normality empirically seems to happen at a much slower rate. This slower rate of convergence is likely also what leads to biased results in the IWOR estimator when all component nuisance models are non-parametrically specified (Table \ref{table:results_ff}), and speaks to the necessity of the influence function based estimator when pursuing a fully flexible approach.
\end{enumerate}

\begin{table}
\centering\footnotesize
\begin{tabularx}{\textwidth}{c@{}c@{}c@{}c@{}c@{}cc@{}c}
\hline
\multicolumn{8}{c}{\textbf{Data Driven Simulation Scenario 1}} \\
\cmidrule{1-8}
\multicolumn{2}{c}{\textbf{Model}} & \textbf{Interactions} & \textbf{Imputation} & \textbf{\% Bias} & \textbf{\% M-Bias} & \textbf{SE} & \textbf{RU}\\
\hline
 & IF &  & --- & 0 & 0 & 0.006 & 1.000\\
\cmidrule{2-2}
\cmidrule{4-8}
\multirow{-2}{*}{\centering\arraybackslash \shortstack{CCMAR-\\based}} & IWOR & \multirow{-2}{*}{\centering\arraybackslash ---} & --- & -1 & -6 & 0.029 & 5.088\\
\cmidrule{1-8}
 &  &  & OLS/LR & 15 & 15 & 0.006 & 1.005\\
\cmidrule{4-8}
 &  &  & OLS/LR w/ Interactions & 12 & 12 & 0.006 & 1.015\\
\cmidrule{4-8}
 &  &  & True DGP & 14 & 14 & 0.006 & 1.005\\
\cmidrule{4-8}
 &  & \multirow{-4}{*}{\centering\arraybackslash None} & --- & -21 & -21 & 0.006 & 1.049\\
\cmidrule{3-8}
 &  &  & OLS/LR & 0 & 0 & 0.005 & 0.812\\
\cmidrule{4-8}
 &  &  & OLS/LR w/ Interactions & 0 & 0 & 0.005 & 0.824\\
\cmidrule{4-8}
 &  &  & True DGP & 0 & 0 & 0.005 & 0.813\\
\cmidrule{4-8}
 & \multirow{-8}{*}{\centering\arraybackslash OLS} & \multirow{-4}{*}{\centering\arraybackslash Pairwise} & --- & -24 & -24 & 0.006 & 0.973\\
\cmidrule{2-8}
 &  &  & OLS/LR & 15 & 15 & 0.006 & 1.007\\
\cmidrule{4-8}
 &  &  & OLS/LR w/ Interactions & 12 & 12 & 0.006 & 1.011\\
\cmidrule{4-8}
 &  &  & True DGP & 14 & 14 & 0.006 & 1.006\\
\cmidrule{4-8}
 &  & \multirow{-4}{*}{\centering\arraybackslash None} & --- & -21 & -21 & 0.006 & 1.050\\
\cmidrule{3-8}
 &  &  & OLS/LR & 0 & 0 & 0.005 & 0.846\\
\cmidrule{4-8}
 &  &  & OLS/LR w/ Interactions & 0 & 0 & 0.005 & 0.848\\
\cmidrule{4-8}
 &  &  & True DGP & 0 & 0 & 0.005 & 0.821\\
\cmidrule{4-8}
 & \multirow{-8}{*}{\centering\arraybackslash GAM} & \multirow{-4}{*}{\centering\arraybackslash Pairwise} & --- & -23 & -23 & 0.006 & 0.982\\
\cmidrule{2-8}
 &  &  & OLS/LR & -1 & -1 & 0.005 & 0.836\\
\cmidrule{4-8}
 &  &  & OLS/LR w/ Interactions & -1 & -1 & 0.005 & 0.835\\
\cmidrule{4-8}
 &  &  & True DGP & -1 & -1 & 0.005 & 0.831\\
\cmidrule{4-8}
\multirow{-20}{*}{\centering\arraybackslash \shortstack{Outcome\\Regression}} & \multirow{-4}{*}{\centering\arraybackslash \shortstack{Random\\Forest}} & \multirow{-4}{*}{\centering\arraybackslash ---} & --- & -30 & -30 & 0.006 & 1.008\\
\cmidrule{1-8}
 &  &  & OLS/LR & 8 & 8 & 0.006 & 0.959\\
\cmidrule{4-8}
 &  &  & OLS/LR w/ Interactions & 6 & 6 & 0.006 & 0.958\\
\cmidrule{4-8}
 &  &  & True DGP & 8 & 8 & 0.006 & 0.965\\
\cmidrule{4-8}
 &  & \multirow{-4}{*}{\centering\arraybackslash None} & --- & -22 & -22 & 0.007 & 1.175\\
\cmidrule{3-8}
 &  &  & OLS/LR & 0 & 0 & 0.005 & 0.857\\
\cmidrule{4-8}
 &  &  & OLS/LR w/ Interactions & 0 & 0 & 0.005 & 0.886\\
\cmidrule{4-8}
 &  &  & True DGP & 1 & 1 & 0.005 & 0.865\\
\cmidrule{4-8}
 & \multirow{-8}{*}{\centering\arraybackslash \shortstack{Logistic\\Regression}} & \multirow{-4}{*}{\centering\arraybackslash Pairwise} & --- & -26 & -26 & 0.007 & 1.203\\
\cmidrule{2-8}
 &  &  & OLS/LR & 6 & 6 & 0.006 & 1.003\\
\cmidrule{4-8}
 &  &  & OLS/LR w/ Interactions & 4 & 4 & 0.006 & 0.988\\
\cmidrule{4-8}
 &  &  & True DGP & 7 & 7 & 0.006 & 0.956\\
\cmidrule{4-8}
 &  & \multirow{-4}{*}{\centering\arraybackslash None} & --- & -23 & -23 & 0.007 & 1.174\\
\cmidrule{3-8}
 &  &  & OLS/LR & 0 & -1 & 0.005 & 0.895\\
\cmidrule{4-8}
 &  &  & OLS/LR w/ Interactions & 0 & 0 & 0.005 & 0.934\\
\cmidrule{4-8}
 &  &  & True DGP & 1 & 1 & 0.005 & 0.839\\
\cmidrule{4-8}
 & \multirow{-8}{*}{\centering\arraybackslash \shortstack{GAM\\(Logit Link)}} & \multirow{-4}{*}{\centering\arraybackslash Pairwise} & --- & -26 & -26 & 0.007 & 1.250\\
\cmidrule{2-8}
 &  &  & OLS/LR & -1 & -1 & 0.005 & 0.896\\
\cmidrule{4-8}
 &  &  & OLS/LR w/ Interactions & -1 & -1 & 0.005 & 0.903\\
\cmidrule{4-8}
 &  &  & True DGP & 1 & 1 & 0.005 & 0.904\\
\cmidrule{4-8}
\multirow{-20}{*}{\centering\arraybackslash IPW} & \multirow{-4}{*}{\centering\arraybackslash \shortstack{Random\\Forest}} & \multirow{-4}{*}{\centering\arraybackslash ---} & --- & -24 & -24 & 0.006 & 0.959\\
\hline
\multicolumn{8}{l}{\rule{0pt}{1em}IF = Influence Function; IWOR = Inverse-Weighted Outcome Regression}\\
\multicolumn{8}{l}{\rule{0pt}{1em}\% Bias/\% M-Bias = 100 $\times$ bias of mean/median ATE estimate relative to true ATE}\\
\multicolumn{8}{l}{\rule{0pt}{1em}Relative Uncertainty (RU) = Ratio of standard error to standard error of CCMAR-based IF Estimator}\\
\multicolumn{8}{l}{\rule{0pt}{1em}Imputation: Linear Regression (OLS), Logistic Regression (LR), True Data Generating Process (DGP)}\\
\end{tabularx}\caption{Data Driven Simulation Scenario \# 1: Factorization in Equation~\eqref{eqn:levis_factorization}, 1 partially missing confounder.}
\label{table:results_1}
\end{table}

\begin{table}
\centering\footnotesize
\begin{tabularx}{\textwidth}{c@{}c@{}c@{}c@{}c@{}cc@{}c}
\hline
\multicolumn{8}{c}{\textbf{Data Driven Simulation Scenario 2}} \\
\cmidrule{1-8}
\multicolumn{2}{c}{\textbf{Model}} & \textbf{Interactions} & \textbf{Imputation} & \textbf{\% Bias} & \textbf{\% M-Bias} & \textbf{SE} & \textbf{RU}\\
\hline
 & IF &  & --- & 0 & 0 & 0.004 & 1.000\\
\cmidrule{2-2}
\cmidrule{4-8}
\multirow{-2}{*}{\centering\arraybackslash \shortstack{CCMAR-\\based}} & IWOR & \multirow{-2}{*}{\centering\arraybackslash ---} & --- & -1 & -5 & 0.020 & 4.587\\
\cmidrule{1-8}
 &  &  & OLS/LR & 5 & 5 & 0.005 & 1.100\\
\cmidrule{4-8}
 &  &  & OLS/LR w/ Interactions & 5 & 5 & 0.005 & 1.099\\
\cmidrule{4-8}
 &  &  & True DGP & 5 & 5 & 0.005 & 1.099\\
\cmidrule{4-8}
 &  & \multirow{-4}{*}{\centering\arraybackslash None} & --- & -23 & -23 & 0.005 & 1.235\\
\cmidrule{3-8}
 &  &  & OLS/LR & 0 & -1 & 0.004 & 0.947\\
\cmidrule{4-8}
 &  &  & OLS/LR w/ Interactions & -1 & -1 & 0.004 & 0.947\\
\cmidrule{4-8}
 &  &  & True DGP & 0 & 0 & 0.004 & 0.948\\
\cmidrule{4-8}
 & \multirow{-8}{*}{\centering\arraybackslash OLS} & \multirow{-4}{*}{\centering\arraybackslash Pairwise} & --- & -19 & -19 & 0.005 & 1.207\\
\cmidrule{2-8}
 &  &  & OLS/LR & 4 & 4 & 0.005 & 1.113\\
\cmidrule{4-8}
 &  &  & OLS/LR w/ Interactions & 4 & 4 & 0.005 & 1.115\\
\cmidrule{4-8}
 &  &  & True DGP & 5 & 5 & 0.005 & 1.100\\
\cmidrule{4-8}
 &  & \multirow{-4}{*}{\centering\arraybackslash None} & --- & -23 & -23 & 0.005 & 1.237\\
\cmidrule{3-8}
 &  &  & OLS/LR & -1 & -1 & 0.004 & 0.989\\
\cmidrule{4-8}
 &  &  & OLS/LR w/ Interactions & -1 & -1 & 0.004 & 0.991\\
\cmidrule{4-8}
 &  &  & True DGP & 0 & 0 & 0.004 & 0.948\\
\cmidrule{4-8}
 & \multirow{-8}{*}{\centering\arraybackslash GAM} & \multirow{-4}{*}{\centering\arraybackslash Pairwise} & --- & -18 & -18 & 0.005 & 1.215\\
\cmidrule{2-8}
 &  &  & OLS/LR & -1 & -1 & 0.004 & 0.974\\
\cmidrule{4-8}
 &  &  & OLS/LR w/ Interactions & -1 & -2 & 0.004 & 0.977\\
\cmidrule{4-8}
 &  &  & True DGP & 0 & 0 & 0.004 & 0.960\\
\cmidrule{4-8}
\multirow{-20}{*}{\centering\arraybackslash \shortstack{Outcome\\Regression}} & \multirow{-4}{*}{\centering\arraybackslash \shortstack{Random\\Forest}} & \multirow{-4}{*}{\centering\arraybackslash ---} & --- & -25 & -25 & 0.006 & 1.267\\
\cmidrule{1-8}
 &  &  & OLS/LR & -4 & -4 & 0.005 & 1.048\\
\cmidrule{4-8}
 &  &  & OLS/LR w/ Interactions & -4 & -4 & 0.005 & 1.046\\
\cmidrule{4-8}
 &  &  & True DGP & -4 & -4 & 0.005 & 1.048\\
\cmidrule{4-8}
 &  & \multirow{-4}{*}{\centering\arraybackslash None} & --- & -26 & -26 & 0.006 & 1.262\\
\cmidrule{3-8}
 &  &  & OLS/LR & 0 & 0 & 0.004 & 0.955\\
\cmidrule{4-8}
 &  &  & OLS/LR w/ Interactions & 0 & 0 & 0.004 & 0.960\\
\cmidrule{4-8}
 &  &  & True DGP & 0 & 0 & 0.004 & 0.956\\
\cmidrule{4-8}
 & \multirow{-8}{*}{\centering\arraybackslash \shortstack{Logistic\\Regression}} & \multirow{-4}{*}{\centering\arraybackslash Pairwise} & --- & -22 & -23 & 0.006 & 1.356\\
\cmidrule{2-8}
 &  &  & OLS/LR & -6 & -6 & 0.006 & 1.320\\
\cmidrule{4-8}
 &  &  & OLS/LR w/ Interactions & -6 & -6 & 0.006 & 1.402\\
\cmidrule{4-8}
 &  &  & True DGP & -4 & -4 & 0.005 & 1.043\\
\cmidrule{4-8}
 &  & \multirow{-4}{*}{\centering\arraybackslash None} & --- & -26 & -26 & 0.006 & 1.267\\
\cmidrule{3-8}
 &  &  & OLS/LR & -2 & -2 & 0.005 & 1.129\\
\cmidrule{4-8}
 &  &  & OLS/LR w/ Interactions & -2 & -2 & 0.005 & 1.177\\
\cmidrule{4-8}
 &  &  & True DGP & 0 & 0 & 0.004 & 0.954\\
\cmidrule{4-8}
 & \multirow{-8}{*}{\centering\arraybackslash \shortstack{GAM\\(Logit Link)}} & \multirow{-4}{*}{\centering\arraybackslash Pairwise} & --- & -22 & -22 & 0.007 & 1.577\\
\cmidrule{2-8}
 &  &  & OLS/LR & -3 & -3 & 0.004 & 0.990\\
\cmidrule{4-8}
 &  &  & OLS/LR w/ Interactions & -3 & -3 & 0.004 & 0.993\\
\cmidrule{4-8}
 &  &  & True DGP & 0 & 0 & 0.004 & 0.984\\
\cmidrule{4-8}
\multirow{-20}{*}{\centering\arraybackslash IPW} & \multirow{-4}{*}{\centering\arraybackslash \shortstack{Random\\Forest}} & \multirow{-4}{*}{\centering\arraybackslash ---} & --- & -21 & -21 & 0.005 & 1.108\\
\hline
\multicolumn{8}{l}{\rule{0pt}{1em}IF = Influence Function; IWOR = Inverse-Weighted Outcome Regression}\\
\multicolumn{8}{l}{\rule{0pt}{1em}\% Bias/\% M-Bias = 100 $\times$ bias of mean/median ATE estimate relative to true ATE}\\
\multicolumn{8}{l}{\rule{0pt}{1em}Relative Uncertainty (RU) = Ratio of standard error to standard error of CCMAR-based IF Estimator}\\
\multicolumn{8}{l}{\rule{0pt}{1em}Imputation: Linear Regression (OLS), Logistic Regression (LR), True Data Generating Process (DGP)}\\
\end{tabularx}\caption{Data Driven Simulation Scenario \# 2: Factorization in Equation~\eqref{eqn:levis_factorization}, 1 partially missing confounder, additional non-linearities and interactions in imputation mean model,induced skew in distribution of imputation model.}
\label{table:results_2}
\end{table}

\begin{table}
\centering\footnotesize
\begin{tabularx}{\textwidth}{c@{}c@{}c@{}c@{}c@{}cc@{}c}
\hline
\multicolumn{8}{c}{\textbf{Data Driven Simulation Scenario 3}} \\
\cmidrule{1-8}
\multicolumn{2}{c}{\textbf{Model}} & \textbf{Interactions} & \textbf{Imputation} & \textbf{\% Bias} & \textbf{\% M-Bias} & \textbf{SE} & \textbf{RU}\\
\hline
 & IF &  & --- & 0 & 0 & 0.007 & 1.000\\
\cmidrule{2-2}
\cmidrule{4-8}
\multirow{-2}{*}{\centering\arraybackslash \shortstack{CCMAR-\\based}} & IWOR & \multirow{-2}{*}{\centering\arraybackslash ---} & --- & -1 & -5 & 0.020 & 2.983\\
\cmidrule{1-8}
 &  &  & OLS/LR & 4 & 4 & 0.005 & 0.759\\
\cmidrule{4-8}
 &  &  & OLS/LR w/ Interactions & 3 & 3 & 0.005 & 0.753\\
\cmidrule{4-8}
 &  &  & True DGP & 3 & 3 & 0.005 & 0.745\\
\cmidrule{4-8}
 &  & \multirow{-4}{*}{\centering\arraybackslash None} & --- & -22 & -23 & 0.006 & 0.825\\
\cmidrule{3-8}
 &  &  & OLS/LR & 0 & 0 & 0.004 & 0.650\\
\cmidrule{4-8}
 &  &  & OLS/LR w/ Interactions & 0 & 0 & 0.004 & 0.646\\
\cmidrule{4-8}
 &  &  & True DGP & 0 & 0 & 0.004 & 0.645\\
\cmidrule{4-8}
 & \multirow{-8}{*}{\centering\arraybackslash OLS} & \multirow{-4}{*}{\centering\arraybackslash Pairwise} & --- & -19 & -19 & 0.005 & 0.804\\
\cmidrule{2-8}
 &  &  & OLS/LR & 3 & 3 & 0.005 & 0.767\\
\cmidrule{4-8}
 &  &  & OLS/LR w/ Interactions & 2 & 2 & 0.005 & 0.769\\
\cmidrule{4-8}
 &  &  & True DGP & 3 & 3 & 0.005 & 0.745\\
\cmidrule{4-8}
 &  & \multirow{-4}{*}{\centering\arraybackslash None} & --- & -22 & -23 & 0.006 & 0.826\\
\cmidrule{3-8}
 &  &  & OLS/LR & -1 & -1 & 0.005 & 0.717\\
\cmidrule{4-8}
 &  &  & OLS/LR w/ Interactions & -1 & -1 & 0.005 & 0.695\\
\cmidrule{4-8}
 &  &  & True DGP & 0 & 0 & 0.004 & 0.644\\
\cmidrule{4-8}
 & \multirow{-8}{*}{\centering\arraybackslash GAM} & \multirow{-4}{*}{\centering\arraybackslash Pairwise} & --- & -18 & -18 & 0.005 & 0.811\\
\cmidrule{2-8}
 &  &  & OLS/LR & -2 & -2 & 0.004 & 0.662\\
\cmidrule{4-8}
 &  &  & OLS/LR w/ Interactions & -2 & -2 & 0.004 & 0.661\\
\cmidrule{4-8}
 &  &  & True DGP & -1 & -1 & 0.004 & 0.650\\
\cmidrule{4-8}
\multirow{-20}{*}{\centering\arraybackslash \shortstack{Outcome\\Regression}} & \multirow{-4}{*}{\centering\arraybackslash \shortstack{Random\\Forest}} & \multirow{-4}{*}{\centering\arraybackslash ---} & --- & -26 & -26 & 0.005 & 0.811\\
\cmidrule{1-8}
 &  &  & OLS/LR & -5 & -5 & 0.005 & 0.744\\
\cmidrule{4-8}
 &  &  & OLS/LR w/ Interactions & -5 & -5 & 0.005 & 0.728\\
\cmidrule{4-8}
 &  &  & True DGP & -5 & -5 & 0.005 & 0.727\\
\cmidrule{4-8}
 &  & \multirow{-4}{*}{\centering\arraybackslash None} & --- & -26 & -26 & 0.006 & 0.893\\
\cmidrule{3-8}
 &  &  & OLS/LR & 1 & 1 & 0.004 & 0.658\\
\cmidrule{4-8}
 &  &  & OLS/LR w/ Interactions & 1 & 1 & 0.004 & 0.658\\
\cmidrule{4-8}
 &  &  & True DGP & 1 & 1 & 0.004 & 0.655\\
\cmidrule{4-8}
 & \multirow{-8}{*}{\centering\arraybackslash \shortstack{Logistic\\Regression}} & \multirow{-4}{*}{\centering\arraybackslash Pairwise} & --- & -22 & -22 & 0.006 & 0.903\\
\cmidrule{2-8}
 &  &  & OLS/LR & -7 & -7 & 0.006 & 0.965\\
\cmidrule{4-8}
 &  &  & OLS/LR w/ Interactions & -7 & -7 & 0.007 & 1.007\\
\cmidrule{4-8}
 &  &  & True DGP & -5 & -5 & 0.005 & 0.723\\
\cmidrule{4-8}
 &  & \multirow{-4}{*}{\centering\arraybackslash None} & --- & -26 & -26 & 0.006 & 0.879\\
\cmidrule{3-8}
 &  &  & OLS/LR & -1 & -1 & 0.005 & 0.818\\
\cmidrule{4-8}
 &  &  & OLS/LR w/ Interactions & -1 & -1 & 0.006 & 0.846\\
\cmidrule{4-8}
 &  &  & True DGP & 1 & 1 & 0.004 & 0.656\\
\cmidrule{4-8}
 & \multirow{-8}{*}{\centering\arraybackslash \shortstack{GAM\\(Logit Link)}} & \multirow{-4}{*}{\centering\arraybackslash Pairwise} & --- & -22 & -22 & 0.007 & 1.035\\
\cmidrule{2-8}
 &  &  & OLS/LR & -2 & -2 & 0.005 & 0.671\\
\cmidrule{4-8}
 &  &  & OLS/LR w/ Interactions & -2 & -2 & 0.004 & 0.668\\
\cmidrule{4-8}
 &  &  & True DGP & 0 & 0 & 0.004 & 0.667\\
\cmidrule{4-8}
\multirow{-20}{*}{\centering\arraybackslash IPW} & \multirow{-4}{*}{\centering\arraybackslash \shortstack{Random\\Forest}} & \multirow{-4}{*}{\centering\arraybackslash ---} & --- & -19 & -19 & 0.005 & 0.747\\
\hline
\multicolumn{8}{l}{\rule{0pt}{1em}IF = Influence Function; IWOR = Inverse-Weighted Outcome Regression}\\
\multicolumn{8}{l}{\rule{0pt}{1em}\% Bias/\% M-Bias = 100 $\times$ bias of mean/median ATE estimate relative to true ATE}\\
\multicolumn{8}{l}{\rule{0pt}{1em}Relative Uncertainty (RU) = Ratio of standard error to standard error of CCMAR-based IF Estimator}\\
\multicolumn{8}{l}{\rule{0pt}{1em}Imputation: Linear Regression (OLS), Logistic Regression (LR), True Data Generating Process (DGP)}\\
\end{tabularx}\caption{Data Driven Simulation Scenario $\# 3$: Factorization in Equation~\eqref{eqn:levis_factorization}, 2 partially missing confounders, amplified interactions between $A, L_{p}$, and $Y$ in joint imputation model, induced skew in distribution of imputation model.}
\label{table:results_3}
\end{table}

\begin{table}
\centering
\footnotesize
\begin{tabularx}{\textwidth}{c@{}c@{}c@{}c@{}c@{}cc@{}c}
\hline
\multicolumn{8}{c}{\textbf{Data Driven Simulation Scenario 4}} \\
\cmidrule{1-8}
\multicolumn{2}{c}{\textbf{Model}} & \textbf{Interactions} & \textbf{Imputation} & \textbf{\% Bias} & \textbf{\% M-Bias} & \textbf{SE} & \textbf{RU}\\
\hline
 & IF &  & --- & 8 & 2 & 0.045 & 1.000\\
\cmidrule{2-2}
\cmidrule{4-8}
\multirow{-2}{*}{\centering\arraybackslash \shortstack{CCMAR-\\based}} & IWOR &  & --- & -2 & -5 & 0.024 & 0.534\\
\cmidrule{1-2}
\cmidrule{4-8}
 & IF &  & --- & 1 & 1 & 0.016 & 0.349\\
\cmidrule{2-2}
\cmidrule{4-8}
\multirow{-2}{*}{\centering\arraybackslash \shortstack{Flexible CCMAR\\(Gamma $\lambda_1$)}} & IWOR &  & --- & -1 & -1 & 0.015 & 0.327\\
\cmidrule{1-2}
\cmidrule{4-8}
 & IF &  & --- & -1 & 1 & 0.061 & 1.363\\
\cmidrule{2-2}
\cmidrule{4-8}
\multirow{-2}{*}{\centering\arraybackslash \shortstack{Flexible CCMAR\\(Guassian $\lambda_1$)}} & IWOR & \multirow{-6}{*}{\centering\arraybackslash ---} & --- & -1 & -2 & 0.015 & 0.328\\
\cmidrule{1-8}
 &  &  & OLS/LR & -1 & -1 & 0.006 & 0.124\\
\cmidrule{4-8}
 &  &  & OLS/LR w/ Interactions & 2 & 2 & 0.007 & 0.151\\
\cmidrule{4-8}
 &  &  & True DGP & 3 & 3 & 0.006 & 0.134\\
\cmidrule{4-8}
 &  & \multirow{-4}{*}{\centering\arraybackslash None} & --- & -12 & -12 & 0.006 & 0.126\\
\cmidrule{3-8}
 &  &  & OLS/LR & -3 & -3 & 0.005 & 0.110\\
\cmidrule{4-8}
 &  &  & OLS/LR w/ Interactions & 0 & 0 & 0.006 & 0.143\\
\cmidrule{4-8}
 &  &  & True DGP & 1 & 1 & 0.005 & 0.113\\
\cmidrule{4-8}
 & \multirow{-8}{*}{\centering\arraybackslash OLS} & \multirow{-4}{*}{\centering\arraybackslash Pairwise} & --- & -9 & -9 & 0.005 & 0.120\\
\cmidrule{2-8}
 &  &  & OLS/LR & -1 & -1 & 0.006 & 0.124\\
\cmidrule{4-8}
 &  &  & OLS/LR w/ Interactions & 2 & 2 & 0.007 & 0.151\\
\cmidrule{4-8}
 &  &  & True DGP & 3 & 3 & 0.006 & 0.134\\
\cmidrule{4-8}
 &  & \multirow{-4}{*}{\centering\arraybackslash None} & --- & -12 & -12 & 0.006 & 0.126\\
\cmidrule{3-8}
 &  &  & OLS/LR & -3 & -3 & 0.005 & 0.111\\
\cmidrule{4-8}
 &  &  & OLS/LR w/ Interactions & 0 & 0 & 0.006 & 0.144\\
\cmidrule{4-8}
 &  &  & True DGP & 2 & 2 & 0.005 & 0.113\\
\cmidrule{4-8}
 & \multirow{-8}{*}{\centering\arraybackslash GAM} & \multirow{-4}{*}{\centering\arraybackslash Pairwise} & --- & -9 & -9 & 0.005 & 0.121\\
\cmidrule{2-8}
 &  &  & OLS/LR & -4 & -4 & 0.005 & 0.111\\
\cmidrule{4-8}
 &  &  & OLS/LR w/ Interactions & -1 & -1 & 0.007 & 0.146\\
\cmidrule{4-8}
 &  &  & True DGP & 1 & 1 & 0.005 & 0.114\\
\cmidrule{4-8}
\multirow{-20}{*}{\centering\arraybackslash \shortstack{Outcome\\Regression}} & \multirow{-4}{*}{\centering\arraybackslash \shortstack{Random\\Forest}} & \multirow{-4}{*}{\centering\arraybackslash ---} & --- & -12 & -12 & 0.005 & 0.120\\
\cmidrule{1-8}
 &  &  & OLS/LR & -4 & -4 & 0.005 & 0.121\\
\cmidrule{4-8}
 &  &  & OLS/LR w/ Interactions & 0 & 0 & 0.007 & 0.158\\
\cmidrule{4-8}
 &  &  & True DGP & 2 & 1 & 0.006 & 0.125\\
\cmidrule{4-8}
 &  & \multirow{-4}{*}{\centering\arraybackslash None} & --- & -12 & -12 & 0.006 & 0.127\\
\cmidrule{3-8}
 &  &  & OLS/LR & -3 & -3 & 0.005 & 0.113\\
\cmidrule{4-8}
 &  &  & OLS/LR w/ Interactions & 1 & 1 & 0.007 & 0.146\\
\cmidrule{4-8}
 &  &  & True DGP & 2 & 2 & 0.005 & 0.119\\
\cmidrule{4-8}
 & \multirow{-8}{*}{\centering\arraybackslash \shortstack{Logistic\\Regression}} & \multirow{-4}{*}{\centering\arraybackslash Pairwise} & --- & -11 & -11 & 0.006 & 0.129\\
\cmidrule{2-8}
 &  &  & OLS/LR & -4 & -4 & 0.006 & 0.128\\
\cmidrule{4-8}
 &  &  & OLS/LR w/ Interactions & -1 & -1 & 0.007 & 0.164\\
\cmidrule{4-8}
 &  &  & True DGP & 1 & 1 & 0.006 & 0.124\\
\cmidrule{4-8}
 &  & \multirow{-4}{*}{\centering\arraybackslash None} & --- & -12 & -12 & 0.006 & 0.127\\
\cmidrule{3-8}
 &  &  & OLS/LR & -3 & -3 & 0.005 & 0.117\\
\cmidrule{4-8}
 &  &  & OLS/LR w/ Interactions & 0 & 0 & 0.007 & 0.148\\
\cmidrule{4-8}
 &  &  & True DGP & 2 & 2 & 0.005 & 0.114\\
\cmidrule{4-8}
 & \multirow{-8}{*}{\centering\arraybackslash \shortstack{GAM\\(Logit Link)}} & \multirow{-4}{*}{\centering\arraybackslash Pairwise} & --- & -10 & -10 & 0.007 & 0.150\\
\cmidrule{2-8}
 &  &  & OLS/LR & -4 & -4 & 0.005 & 0.114\\
\cmidrule{4-8}
 &  &  & OLS/LR w/ Interactions & -3 & -3 & 0.006 & 0.127\\
\cmidrule{4-8}
 &  &  & True DGP & -2 & -2 & 0.005 & 0.121\\
\cmidrule{4-8}
\multirow{-20}{*}{\centering\arraybackslash IPW} & \multirow{-4}{*}{\centering\arraybackslash \shortstack{Random\\Forest}} & \multirow{-4}{*}{\centering\arraybackslash ---} & --- & -12 & -12 & 0.005 & 0.119\\
\hline
\multicolumn{8}{l}{\rule{0pt}{1em}IF = Influence Function; IWOR = Inverse-Weighted Outcome Regression}\\
\multicolumn{8}{l}{\rule{0pt}{1em}\% Bias/\% M-Bias = 100 $\times$ bias of mean/median ATE estimate relative to true ATE}\\
\multicolumn{8}{l}{\rule{0pt}{1em}Relative Uncertainty (RU) = Ratio of standard error to standard error of CCMAR-based IF Estimator}\\
\multicolumn{8}{l}{\rule{0pt}{1em}Imputation: Linear Regression (OLS), Logistic Regression (LR), True Data Generating Process (DGP)}\\
\multicolumn{8}{l}{\rule{0pt}{1em}Simulations dropped due to extreme results (out of 5,000): $^*$ (2), $^\dagger$ (7), $\ddagger$ (96)}
\end{tabularx}\caption{Data Driven Simulation Scenario \#4: Alternative factorization, 2 partially missing confounders, amplified interactions between $L_{c}, L_{p}$ in full treatment model $\tilde{\eta}(L_c, L_p, a)$.}
\label{table:results_4}
\end{table}

\clearpage

\begin{table}
\centering
\begin{tabular}{ScScccc@{}c}
\hline
\multicolumn{6}{c}{\textbf{Non-Parametric Nuisance Model Simulation}} \\
\cmidrule{1-6}
\textbf{Estimator} & \textbf{Nusiance Models} & \textbf{\% Bias} & \textbf{\% M-Bias} & \textbf{SE} & \textbf{RU}\\
\hline
 & Parametric & 0.6 & 0.6 & 0.009 & 1.000\\
\cmidrule{2-6}
\multirow{-2}{*}{\centering\arraybackslash CCMAR IF} & Non-Parametric & 0.6 & 1.1 & 0.013 & 1.390\\
\cmidrule{1-6}
 & Parametric & 0.7 & 0.9 & 0.007 & 0.734\\
\cmidrule{2-6}
\multirow{-2}{*}{\centering\arraybackslash CCMAR IWOR} & Non-Parametric & 5.9 & 5.9 & 0.008 & 0.865\\
\hline
\multicolumn{6}{l}{\rule{0pt}{1em}IF = Influence Function; IWOR = Inverse-Weighted Outcome Regression}\\
\multicolumn{6}{l}{\rule{0pt}{1em}\% Bias/\% M-Bias = 100 $\times$ bias of mean/median ATE estimate relative to true ATE}\\
\multicolumn{6}{L{\textwidth}}{\rule{0pt}{1em}Relative Uncertainty (RU) = Ratio of standard error to standard error of parametric CCMAR-based influence function estimator}\\
\end{tabular}\caption{Results from simulation utilizing non-parametric estimation procedures for component nuisance functions in Levis estimators.}
\label{table:results_ff}
\end{table}

\section{Discussion}\label{sec:discussion}
It is inevitable that any analyst seeking to use electronic heath records, or other observational databases for that matter, will have to contend with the simultaneous challenges of missing data and confounding. In standard practice, analysts choose to address these challenges separately, through some combination of imputation to deal with missing data along with standard causal inference methods to deal with confounding. Whether or not such analysis techniques will perform well certainly depends on the setting. Furthermore, it is often the case that the complexity and breadth of assumptions invoked when conducting such \textit{ad hoc} approaches are not clear, nor whether approaches perform well in terms of efficiency and/or robustness to model misspecification.

Understanding that no simulation study can hope to cover all possible scenarios, we believe that those that we do consider collectively provide useful insight on key decisions that analysts have to engage in. Furthermore, we believe that the framework we present provides a path for researchers designing their own related simulations, perhaps tailored to settings they find themselves in. Our code is made publicly available at \url{https://github.com/lbenz730/missing_confounder_sims/}, and we hope researchers interested in additional scenarios or other alternatives can use our resources as a starting point to run their own simulations.

Overall, our results should offer encouraging evidence that reasonable choices, in the form of a suite of \textit{ad hoc} combinations of standard methods for imputation and causal inference, are indeed often reasonable. Across several simulation scenarios, some subset of the many \textit{ad hoc} methods had small biases and in many cases, were more efficient than $\hat\chi_1 - \hat\chi_0$, the influence function CCMAR-based estimator of Levis \etal\cite{levis2022}. Given the relative ease of implementation of these estimators, their use may be well justified in many common situations.

Unfortunately, not every method performed well in every scenario. This is not necessarily surprising, and it seems unrealistic that one method will always outperform all other methods for every problem. While there are numerous options analysts have for dealing with partially missing confounders, we do not believe in, nor advocate for, a one-method fits all approach. Instead, we encourage careful consideration of bias, efficiency, and robustness, and believe that CCMAR-IF estimator of Levis \etal \cite{levis2022} is reasonable, pragmatic choice if analysts want to be cautious, particularly when little is known about underlying treatment, outcome, and/or missingness mechanisms. Of course the CCMAR-IF estimator is not without its own limitations, and is unlikely to be a practical choice when $L_p$ is high dimensional. Though improved methods for conditional density estimation (both statistically and computationally) will be developed as time goes on \citep{hejazi2022haldensify-joss, hejazi2022haldensify-rpkg}, density estimation remains a computational bottleneck for the CCMAR-based estimators at present. 

Though we used single imputation rather than multiple imputation to lower the computational burden of our simulation study, we anticipate similar conclusions from using multiple imputation. Since our imputation scheme is equivalent to multiple imputation with $M = 1$, increasing the number of imputed datasets will not change bias properties of the estimators considered, though the variability of estimators across simulated datasets may be slightly smaller. Because non-trivial bias remained in many scenarios even when missing data were imputed by the true data generating process, more complex imputation methods are not a remedy for misspecification of models necessary to adjust for confounding. Nevertheless, methods like multiple imputation by chained equations (MICE) \cite{vanBuuren2011}, joint modeling multiple imputation (JOMO) \cite{Quartagno2016}, and substantive model compatible fully conditional specification (SMCFCS) \cite{Bartlett2015} may be useful tools for consideration in more complex settings when $L_p$ is high dimensional or missingness patterns are non-monotone. Practically, however, state-of-the-art methods for non-monotone MAR data \cite{sun2018} rely on a set of parametric models for the missingness probabilities, which unfortunately involve unobservable strata. Additional efficiency considerations based on this direction are discussed in more detail in Levis et al. \cite{levis2022}, but such discussion is outside of the primary scope of our work.

Given the lack of a uniformly best solution, we suggest some additional guiding principles for analysts faced with the problem of simultaneous missing data and confounding.

\begin{itemize}
    \item \textbf{Do not ignore missing data}: While dropping subjects with incomplete data offers the path of least resistance, invoking the assumption that confounders are missing completely at random is unlikely to hold in practice, at least in EHR-based observational studies. Across simulation settings, complete case analyses led to both efficiency loss and biased estimates of causal contrasts of interest. Meanwhile, standard causal methods performed reasonably well even when the imputation model was misspecified to a degree, suggesting one needn't let lack of knowledge about the exact imputation model be a deterrent. Imputation is thus a critical tool in any analyst's tool box when tackling this problem. 
    \item \textbf{Do not be afraid of flexible modeling choices}: Using semi-parametric (GAM) or non-parametric (random forest) methods to model component nuisance functions was generally a good strategy. While one might think that in general, flexible modeling choices come at the expense of efficiency losses, our results suggest that may not always be the case. While such models may be slightly more difficult to use or interpret, we think their improvement in estimating causal contrasts of interest makes them worth pursing. 
    \item \textbf{Take care in specifying complex models}: Though complex models generally yielded reduced bias in estimating the average treatment effect, employing such methods often required greater adjustment of default settings compared to simpler parametric models. Analysts should take care in specifying interactions, especially when using GAMs, and consider regularization methods to help select which specified interactions are truly important \cite{hernan2020}. Random forest parameters, such as tree depth, number of component trees, or the number of randomly drawn candidate variables considered at each split, should ideally be chosen based on some cross-validation strategy rather than simply using default algorithm settings. Even when using parametric models, distributional assumptions (perhaps in the form of link function) matter. As in other settings, analysts should take care to consider subject matter expertise and past research when performing causal inference, but equal care must be given towards good modeling principles.
\end{itemize}

EHR databases will only grow in size and promise in years to come, and so too will attempts to use such observational data for causal inference. Given how common an analysis plan of $\textrm{Imputation} +\{\textrm{Outcome Regression}, \textrm{Inverse Probability Weighting}\}$ is, we are glad to know that past analysts who have been principled in their approaches have conducted causal inference that seems both reasonable and valid, and so too will future analysts who adhere to the principles outlined in this paper.

\clearpage
\section*{Declarations}
\subsection*{Ethics approval and consent to participate}
Not applicable

\subsection*{Consent for publication}
Not applicable 

\subsection*{Availability of data and materials}
Additional information on simulation scenarios \#5-19 can be found in our Supplementary Materials online. Code can be found online at \url{https://github.com/lbenz730/missing_confounder_sims}.

\subsection*{Competing Interests}
Not applicable

\subsection*{Funding}
This work is funded by NIH grants R01 DK128150-01 (LB, SH) and 1F31DK141237-01 (LB).

\subsection*{Authors' contributions}
LB and AL designed the experiments and implemented code for simulation studies. AL and SH oversaw project management of the study. All authors wrote and reviewed the manuscript.

\subsection*{Acknowledgments}
Not applicable


\bibliography{bibliography}

\end{document}


\title{Supplementary Materials for ``Comparing Causal Inference Methods for Point Exposures with Missing Confounders: A Simulation Study''}


\author*[1]{\fnm{Luke} \sur{Benz}}\email{lukebenz@g.harvard.edu}
\author[2]{\fnm{Alexander W.} \sur{Levis}}
\author[1]{\fnm{Sebastien} \sur{Haneuse}}

\affil[1]{\orgdiv{Department of Biostatistics}, \orgname{Harvard T.H. Chan School of Public Health}, \orgaddress{\state{Massachusetts}, \country{U.S.A}}}

\affil[2]{\orgdiv{Department of Statistics \& Data Science}, \orgname{Carnegie Mellon University}, \orgaddress{\state{Pennsylvania}, \country{U.S.A}}}

\maketitle

\section{Additional Simulation Scenarios}\label{sec:sim_scenarios}
In the main body of our paper, we presented the results of 4 data-driven simulation scenarios that best summarized our findings after analyzing the estimators of Levis \etal\cite{levis2022} along with a suite of reasonable ad-hoc approaches across a total of 19 simulation settings. Details on the additional 15 simulation settings not presented in the main paper are shown here. The additional simulation scenarios considered were as follows. Note that references to equations refer to their numbering in the main body of the paper.

\begin{enumerate}
    \setcounter{enumi}{4}
    \item{Replication of simulation from Levis \etal\cite{levis2022} paper: Factorization in Equation (1), 1 partially missing confounder.}
    \item{Factorization in Equation (1), 1 partially missing confounder, dampen some of the effects amplified in treatment model and imputation model in scenario \#5 to correct near-positivity violations.}
    \item{Factorization in Equation (1), 1 partially missing confounder, amplify interactions in treatment model.}
    \item{Factorization in Equation (1), 1 partially missing confounder, introduce non-linearities in imputation model.}
    \item{Factorization in Equation (1), 2 partially missing confounders, extend scenario \#6 to have a second partially missing confounder.}
    \item{Factorization in Equation (1), 2 partially missing confounders, extend scenario \#7 to have a second partially missing confounder.}
    \item{Factorization in Equation (1), 2 partially missing confounders, extend scenario \#1 to have a second partially missing confounder (baseline simulation in main paper with a second missing confounder).}
    \item{Factorization in Equation (1), 2 partially missing confounder, extend scenario \#8 to have a second partially missing confounder.}
    \item{Factorization in Equation (1), 2 partially missing confounder, extend scenario \#2 to have a second partially missing confounder.}
    \item{Factorization in Equation (1), 2 partially missing confounders, additional non-linearities and interactions in imputation model for first missing confounder, induce skew in imputation model for first missing confounder, imputation model for second missing confounder depends only on $L_{p1}$ with amplified relationship to emphasize importance of modeling $\lambda_1(\ell_{p1} ~|~ L_c, A, Y, S = 1)$ correctly.}
    \item{Factorization in Equation (1), 2 partially missing confounders, extend scenario \#14 such that imputation model for second missing confounder depends only on both $L_{p1}$ and $Y$.}
    \item{Factorization in Equation (1),, 2 partially missing confounders amplify effect of outcome $Y$ in imputation model for second missing confounder.}
    \item{Factorization in Equation (1), 2 partially missing confounders, introduce interaction between treatment $A$ and outcome $Y$ in imputation model for second missing confounder.}
    \item{Alternative factorization, 1 partially missing confounder.}
    \item{Alternative factorization, 1 partially missing confounder, introduce interactions between $L_c$ and $L_p$ in treatment model $\tilde\eta(L_c, L_p a)$.}
\end{enumerate}

\subsection{Simulation Parameters}\label{sec:sim_params}
Covariates and coefficients for each model from simulation scenarios 5-19 are presented in Tables \ref{tab:coefficients_sup1} - \ref{tab:coefficients_sup4}

\begin{sidewaystable}
    \footnotesize
\begin{minipage}[t]{0.5\textwidth}
\centering

\begin{tabular}[t]{>{}ccccc>{}c}
\hline
\multicolumn{1}{c}{\textbf{ }} & \multicolumn{1}{c}{\textbf{ }} & \multicolumn{4}{c}{\textbf{Data Driven Simulation Scenario}} \\
\cmidrule{3-6}
\textbf{Model} & \textbf{Term} & \textbf{5} & \textbf{6} & \textbf{7} & \textbf{8}\\
\hline
 & (Intercept) & -0.624 & -0.624 & -0.624 & -0.624\\
\cmidrule{2-6}
 & $L^{(1)}$ & 0.508 & 0.308 & 0.308 & 0.308\\
\cmidrule{2-6}
 & $L^{(2)}$ & -0.046 & -0.046 & -0.046 & -0.046\\
\cmidrule{2-6}
 & $L^{(3)}$ & 0.435 & 0.385 & 0.385 & 0.385\\
\cmidrule{2-6}
 & $L^{(4)}$ & --- & --- & --- & ---\\
\cmidrule{2-6}
 & $(L^{(2)})^2$ & 0.001 & 0.001 & 0.001 & 0.001\\
\cmidrule{2-6}
 & $L^{(1)} \times L^{(3)}$ & 0.400 & 0.100 & 0.400 & 0.400\\
\cmidrule{2-6}
\multirow{-8}{*}{\centering\arraybackslash $\eta~\text{or}~\tilde\eta$} & $L^{(3)} \times L^{(4)}$ & --- & --- & --- & ---\\
\cmidrule{1-6}
 & (Intercept) & -0.207 & -0.207 & -0.207 & -0.207\\
\cmidrule{2-6}
 & $L^{(1)}$ & 0.031 & 0.031 & 0.031 & 0.031\\
\cmidrule{2-6}
 & $L^{(2)}$ & -0.002 & -0.002 & -0.002 & -0.002\\
\cmidrule{2-6}
 & $L^{(3)}$ & 0.023 & 0.023 & 0.023 & 0.023\\
\cmidrule{2-6}
 & $L^{(4)}$ & --- & --- & --- & ---\\
\cmidrule{2-6}
 & $L^{(5)}$ & --- & --- & --- & ---\\
\cmidrule{2-6}
 & $L^{(1)} \times L^{(3)}$ & -0.305 & -0.305 & -0.305 & -0.305\\
\cmidrule{2-6}
 & $L^{(4)} \times L^{(5)}$ & --- & --- & --- & ---\\
\cmidrule{2-6}
 & $A$ & 0.045 & 0.045 & 0.045 & 0.045\\
\cmidrule{2-6}
 & $A \times L^{(1)}$ & 0.313 & 0.313 & 0.313 & 0.313\\
\cmidrule{2-6}
 & $A \times L^{(2)}$ & -0.001 & -0.001 & -0.001 & -0.001\\
\cmidrule{2-6}
 & $A \times L^{(3)}$ & 0.080 & 0.080 & 0.080 & 0.080\\
\cmidrule{2-6}
 & $A \times L^{(4)}$ & --- & --- & --- & ---\\
\cmidrule{2-6}
 & $A \times L^{(5)}$ & --- & --- & --- & ---\\
\cmidrule{2-6}
\multirow{-15}{*}{\centering\arraybackslash $\mu~\text{or}~\tilde\mu$} & $\sigma$ & 0.109 & 0.109 & 0.109 & 0.109\\
\cmidrule{1-6}
 & (Intercept) & 1.824 & 1.824 & 1.824 & 1.824\\
\cmidrule{2-6}
 & $L^{(1)}$ & 0.087 & 0.087 & 0.087 & 0.087\\
\cmidrule{2-6}
 & $L^{(2)}$ & 0.010 & 0.010 & 0.010 & 0.010\\
\cmidrule{2-6}
 & $L^{(3)}$ & -2.662 & -2.662 & -2.662 & -2.662\\
\cmidrule{2-6}
 & $L^{(2)} \times L^{(3)}$ & -0.149 & -0.149 & -0.149 & -0.149\\
\cmidrule{2-6}
 & $A$ & 2.922 & 2.922 & 2.922 & 2.922\\
\cmidrule{2-6}
 & $Y$ & 2.180 & 2.180 & 2.180 & 2.180\\
\cmidrule{2-6}
 & $A\times Y$ & 3.043 & 3.043 & 3.043 & 3.043\\
\cmidrule{2-6}
 & $A \times L^{(2)}$ & 0.159 & 0.159 & 0.159 & 0.159\\
\cmidrule{2-6}
\multirow{-10}{*}{\centering\arraybackslash $\pi$} & $Y \times L^{(1)}$ & 2.321 & 2.321 & 2.321 & 2.321\\
\hline
\end{tabular}
\end{minipage}
\begin{minipage}[t]{0.5\textwidth}

\begin{tabular}[t]{>{}ccccc>{}c}
\hline
\multicolumn{1}{c}{\textbf{ }} & \multicolumn{1}{c}{\textbf{ }} & \multicolumn{4}{c}{\textbf{Data Driven Simulation Scenario}} \\
\cmidrule{3-6}
\textbf{Model} & \textbf{Term} & \textbf{5} & \textbf{6} & \textbf{7} & \textbf{8}\\
\hline
 & (Intercept) & 0.867 & 0.867 & 0.867 & 0.939\\
\cmidrule{2-6}
 & $L^{(1)}$ & 0.075 & 0.075 & 0.075 & 0.074\\
\cmidrule{2-6}
 & $L^{(2)}$ & < 0.001 & < 0.001 & < 0.001 & -0.018\\
\cmidrule{2-6}
 & $L^{(3)}$ & -0.034 & -0.034 & -0.034 & -0.033\\
\cmidrule{2-6}
 & $(L^{(2)})^2$ & --- & --- & --- & < 0.001\\
\cmidrule{2-6}
 & $L^{(1)} \times L^{(2)}$ & --- & --- & --- & ---\\
\cmidrule{2-6}
 & $L^{(1)} \times L^{(3)}$ & --- & --- & --- & ---\\
\cmidrule{2-6}
 & $A$ & -0.597 & 0.103 & 0.103 & 0.102\\
\cmidrule{2-6}
 & $Y$ & -1.465 & -0.765 & -0.765 & -0.696\\
\cmidrule{2-6}
 & $A\times Y$ & -0.400 & -0.050 & -0.050 & -0.050\\
\cmidrule{2-6}
 & $Y \times L^{(2)}$ & --- & --- & --- & -0.005\\
\cmidrule{2-6}
\multirow{-12}{*}{\centering\arraybackslash $\lambda_1~\text{or}~\tilde\lambda_1$} & $\alpha$ & 3.619 & 3.619 & 3.619 & 3.622\\
\cmidrule{1-6}
 & (Intercept) & --- & --- & --- & ---\\
\cmidrule{2-6}
 & $L^{(1)}$ & --- & --- & --- & ---\\
\cmidrule{2-6}
 & $L^{(2)}$ & --- & --- & --- & ---\\
\cmidrule{2-6}
 & $L^{(3)}$ & --- & --- & --- & ---\\
\cmidrule{2-6}
 & $L^{(4)}$ & --- & --- & --- & ---\\
\cmidrule{2-6}
 & $A$ & --- & --- & --- & ---\\
\cmidrule{2-6}
 & $Y$ & --- & --- & --- & ---\\
\cmidrule{2-6}
 & $A\times Y$ & --- & --- & --- & ---\\
\cmidrule{2-6}
 & $A \times L^{(4)}$ & --- & --- & --- & ---\\
\cmidrule{2-6}
 & $Y \times L^{(2)}$ & --- & --- & --- & ---\\
\cmidrule{2-6}
\multirow{-11}{*}{\centering\arraybackslash $\lambda_2~\text{or}~\tilde\lambda_2$} & $Y \times L^{(4)}$ & --- & --- & --- & ---\\
\hline
\end{tabular}
\end{minipage}

\caption{Coefficients ($\beta$) and covariates ($X$) used to generated simulated datasets for simulation scenarios 5-8. All four scenarios generated datasets using the factorization from Equation (1). --- denotes the given covariate was not including in the corresponding model. Note that all 4 scenarios only generated a single partially missing confounder and as such no $\lambda_2$ model was needed.}
\label{tab:coefficients_sup1}
\end{sidewaystable}

\begin{sidewaystable}
    \footnotesize
\begin{minipage}[t]{0.5\textwidth}
\centering

\begin{tabular}[t]{>{}ccccc>{}c}
\hline
\multicolumn{1}{c}{\textbf{ }} & \multicolumn{1}{c}{\textbf{ }} & \multicolumn{4}{c}{\textbf{Data Driven Simulation Scenario}} \\
\cmidrule{3-6}
\textbf{Model} & \textbf{Term} & \textbf{9} & \textbf{10} & \textbf{11} & \textbf{12}\\
\hline
 & (Intercept) & -0.624 & -0.624 & -0.624 & -0.624\\
\cmidrule{2-6}
 & $L^{(1)}$ & 0.308 & 0.308 & 0.308 & 0.308\\
\cmidrule{2-6}
 & $L^{(2)}$ & -0.046 & -0.046 & -0.046 & -0.046\\
\cmidrule{2-6}
 & $L^{(3)}$ & 0.385 & 0.385 & -0.015 & 0.385\\
\cmidrule{2-6}
 & $L^{(4)}$ & --- & --- & --- & ---\\
\cmidrule{2-6}
 & $(L^{(2)})^2$ & 0.001 & 0.001 & 0.001 & 0.001\\
\cmidrule{2-6}
 & $L^{(1)} \times L^{(3)}$ & 0.100 & 0.400 & 0.400 & 0.400\\
\cmidrule{2-6}
\multirow{-8}{*}{\centering\arraybackslash $\eta~\text{or}~\tilde\eta$} & $L^{(3)} \times L^{(4)}$ & --- & --- & --- & ---\\
\cmidrule{1-6}
 & (Intercept) & -0.207 & -0.207 & -0.207 & -0.207\\
\cmidrule{2-6}
 & $L^{(1)}$ & 0.031 & 0.031 & 0.031 & 0.031\\
\cmidrule{2-6}
 & $L^{(2)}$ & -0.002 & -0.002 & -0.002 & -0.002\\
\cmidrule{2-6}
 & $L^{(3)}$ & 0.023 & 0.023 & 0.023 & 0.023\\
\cmidrule{2-6}
 & $L^{(4)}$ & --- & --- & --- & ---\\
\cmidrule{2-6}
 & $L^{(5)}$ & --- & --- & --- & ---\\
\cmidrule{2-6}
 & $L^{(1)} \times L^{(3)}$ & -0.305 & -0.305 & -0.305 & -0.305\\
\cmidrule{2-6}
 & $L^{(4)} \times L^{(5)}$ & --- & --- & --- & ---\\
\cmidrule{2-6}
 & $A$ & 0.045 & 0.045 & 0.045 & 0.045\\
\cmidrule{2-6}
 & $A \times L^{(1)}$ & 0.313 & 0.313 & 0.413 & 0.313\\
\cmidrule{2-6}
 & $A \times L^{(2)}$ & -0.001 & -0.001 & -0.003 & -0.001\\
\cmidrule{2-6}
 & $A \times L^{(3)}$ & 0.080 & 0.080 & 0.130 & 0.080\\
\cmidrule{2-6}
 & $A \times L^{(4)}$ & --- & --- & --- & ---\\
\cmidrule{2-6}
 & $A \times L^{(5)}$ & --- & --- & --- & ---\\
\cmidrule{2-6}
\multirow{-15}{*}{\centering\arraybackslash $\mu~\text{or}~\tilde\mu$} & $\sigma$ & 0.109 & 0.109 & 0.109 & 0.109\\
\cmidrule{1-6}
 & (Intercept) & 1.824 & 1.824 & 1.824 & 1.824\\
\cmidrule{2-6}
 & $L^{(1)}$ & 0.087 & 0.087 & 0.087 & 0.087\\
\cmidrule{2-6}
 & $L^{(2)}$ & 0.010 & 0.010 & 0.010 & 0.010\\
\cmidrule{2-6}
 & $L^{(3)}$ & -2.662 & -2.662 & -2.662 & -2.662\\
\cmidrule{2-6}
 & $L^{(2)} \times L^{(3)}$ & -0.149 & -0.149 & -0.149 & -0.149\\
\cmidrule{2-6}
 & $A$ & 2.922 & 2.922 & 2.922 & 2.922\\
\cmidrule{2-6}
 & $Y$ & 2.180 & 2.180 & 2.180 & 2.180\\
\cmidrule{2-6}
 & $A\times Y$ & 3.043 & 3.043 & 3.043 & 3.043\\
\cmidrule{2-6}
 & $A \times L^{(2)}$ & 0.159 & 0.159 & 0.159 & 0.159\\
\cmidrule{2-6}
\multirow{-10}{*}{\centering\arraybackslash $\pi$} & $Y \times L^{(1)}$ & 2.321 & 2.321 & 2.321 & 2.321\\
\hline
\end{tabular}
\end{minipage}
\begin{minipage}[t]{0.5\textwidth}

\begin{tabular}[t]{>{}ccccc>{}c}
\hline
\multicolumn{1}{c}{\textbf{ }} & \multicolumn{1}{c}{\textbf{ }} & \multicolumn{4}{c}{\textbf{Data Driven Simulation Scenario}} \\
\cmidrule{3-6}
\textbf{Model} & \textbf{Term} & \textbf{9} & \textbf{10} & \textbf{11} & \textbf{12}\\
\hline
 & (Intercept) & 0.867 & 0.867 & 0.867 & 0.939\\
\cmidrule{2-6}
 & $L^{(1)}$ & 0.075 & 0.075 & 0.075 & 0.074\\
\cmidrule{2-6}
 & $L^{(2)}$ & < 0.001 & < 0.001 & < 0.001 & -0.018\\
\cmidrule{2-6}
 & $L^{(3)}$ & -0.034 & -0.034 & -0.034 & -0.033\\
\cmidrule{2-6}
 & $(L^{(2)})^2$ & --- & --- & --- & < 0.001\\
\cmidrule{2-6}
 & $L^{(1)} \times L^{(2)}$ & --- & --- & --- & ---\\
\cmidrule{2-6}
 & $L^{(1)} \times L^{(3)}$ & --- & --- & --- & ---\\
\cmidrule{2-6}
 & $A$ & 0.103 & 0.103 & 0.303 & 0.102\\
\cmidrule{2-6}
 & $Y$ & -0.765 & -0.765 & -0.765 & -0.696\\
\cmidrule{2-6}
 & $A\times Y$ & -0.050 & -0.050 & -0.500 & -0.050\\
\cmidrule{2-6}
 & $Y \times L^{(2)}$ & --- & --- & --- & -0.005\\
\cmidrule{2-6}
\multirow{-12}{*}{\centering\arraybackslash $\lambda_1~\text{or}~\tilde\lambda_1$} & $\alpha$ & 3.619 & 3.619 & 3.619 & 3.622\\
\cmidrule{1-6}
 & (Intercept) & -1.386 & -1.386 & -1.386 & -1.386\\
\cmidrule{2-6}
 & $L^{(1)}$ & 0.070 & 0.070 & 0.070 & 0.070\\
\cmidrule{2-6}
 & $L^{(2)}$ & 0.040 & 0.040 & 0.040 & 0.040\\
\cmidrule{2-6}
 & $L^{(3)}$ & 0.050 & 0.050 & 0.050 & 0.050\\
\cmidrule{2-6}
 & $L^{(4)}$ & 0.050 & 0.050 & 0.050 & 0.050\\
\cmidrule{2-6}
 & $A$ & 0.030 & 0.030 & 0.030 & 0.030\\
\cmidrule{2-6}
 & $Y$ & 0.250 & 0.250 & 0.250 & 0.250\\
\cmidrule{2-6}
 & $A\times Y$ & --- & --- & --- & ---\\
\cmidrule{2-6}
 & $A \times L^{(4)}$ & --- & --- & --- & ---\\
\cmidrule{2-6}
 & $Y \times L^{(2)}$ & 0.050 & 0.050 & 0.050 & 0.050\\
\cmidrule{2-6}
\multirow{-11}{*}{\centering\arraybackslash $\lambda_2~\text{or}~\tilde\lambda_2$} & $Y \times L^{(4)}$ & --- & --- & --- & ---\\
\hline
\end{tabular}
\end{minipage}

\caption{Coefficients ($\beta$) and covariates ($X$) used to generated simulated datasets for simulation scenarios 9-12. All four scenarios generated datasets using the factorization from Equation (1). --- denotes the given covariate was not including in the corresponding model.}
\label{tab:coefficients_sup2}
\end{sidewaystable}

\begin{sidewaystable}
    \footnotesize
\begin{minipage}[t]{0.5\textwidth}
\centering

\begin{tabular}[t]{>{}ccccc>{}c}
\hline
\multicolumn{1}{c}{\textbf{ }} & \multicolumn{1}{c}{\textbf{ }} & \multicolumn{4}{c}{\textbf{Data Driven Simulation Scenario}} \\
\cmidrule{3-6}
\textbf{Model} & \textbf{Term} & \textbf{13} & \textbf{14} & \textbf{15} & \textbf{16}\\
\hline
 & (Intercept) & -0.624 & -0.624 & -0.624 & -0.624\\
\cmidrule{2-6}
 & $L^{(1)}$ & 0.308 & 0.308 & 0.308 & 0.308\\
\cmidrule{2-6}
 & $L^{(2)}$ & -0.046 & -0.046 & -0.046 & -0.046\\
\cmidrule{2-6}
 & $L^{(3)}$ & 0.385 & 0.385 & 0.385 & 0.385\\
\cmidrule{2-6}
 & $L^{(4)}$ & --- & --- & --- & ---\\
\cmidrule{2-6}
 & $(L^{(2)})^2$ & 0.001 & 0.001 & 0.001 & 0.001\\
\cmidrule{2-6}
 & $L^{(1)} \times L^{(3)}$ & 0.400 & 0.400 & 0.400 & 0.400\\
\cmidrule{2-6}
\multirow{-8}{*}{\centering\arraybackslash $\eta~\text{or}~\tilde\eta$} & $L^{(3)} \times L^{(4)}$ & --- & --- & --- & ---\\
\cmidrule{1-6}
 & (Intercept) & -0.207 & -0.207 & -0.207 & -0.207\\
\cmidrule{2-6}
 & $L^{(1)}$ & 0.031 & 0.031 & 0.031 & 0.031\\
\cmidrule{2-6}
 & $L^{(2)}$ & -0.002 & -0.002 & -0.002 & -0.002\\
\cmidrule{2-6}
 & $L^{(3)}$ & 0.023 & 0.023 & 0.023 & 0.023\\
\cmidrule{2-6}
 & $L^{(4)}$ & --- & --- & --- & ---\\
\cmidrule{2-6}
 & $L^{(5)}$ & --- & --- & --- & ---\\
\cmidrule{2-6}
 & $L^{(1)} \times L^{(3)}$ & -0.305 & -0.305 & -0.305 & -0.305\\
\cmidrule{2-6}
 & $L^{(4)} \times L^{(5)}$ & --- & --- & --- & ---\\
\cmidrule{2-6}
 & $A$ & 0.045 & 0.045 & 0.045 & 0.045\\
\cmidrule{2-6}
 & $A \times L^{(1)}$ & 0.313 & 0.313 & 0.313 & 0.313\\
\cmidrule{2-6}
 & $A \times L^{(2)}$ & -0.001 & -0.001 & -0.001 & -0.001\\
\cmidrule{2-6}
 & $A \times L^{(3)}$ & 0.080 & 0.080 & 0.080 & 0.080\\
\cmidrule{2-6}
 & $A \times L^{(4)}$ & --- & --- & --- & ---\\
\cmidrule{2-6}
 & $A \times L^{(5)}$ & --- & --- & --- & ---\\
\cmidrule{2-6}
\multirow{-15}{*}{\centering\arraybackslash $\mu~\text{or}~\tilde\mu$} & $\sigma$ & 0.109 & 0.109 & 0.109 & 0.109\\
\cmidrule{1-6}
 & (Intercept) & 1.824 & 1.824 & 1.824 & 1.824\\
\cmidrule{2-6}
 & $L^{(1)}$ & 0.087 & 0.087 & 0.087 & 0.087\\
\cmidrule{2-6}
 & $L^{(2)}$ & 0.010 & 0.010 & 0.010 & 0.010\\
\cmidrule{2-6}
 & $L^{(3)}$ & -2.662 & -2.662 & -2.662 & -2.662\\
\cmidrule{2-6}
 & $L^{(2)} \times L^{(3)}$ & -0.149 & -0.149 & -0.149 & -0.149\\
\cmidrule{2-6}
 & $A$ & 2.922 & 2.922 & 2.922 & 2.922\\
\cmidrule{2-6}
 & $Y$ & 2.180 & 2.180 & 2.180 & 2.180\\
\cmidrule{2-6}
 & $A\times Y$ & 3.043 & 3.043 & 3.043 & 3.043\\
\cmidrule{2-6}
 & $A \times L^{(2)}$ & 0.159 & 0.159 & 0.159 & 0.159\\
\cmidrule{2-6}
\multirow{-10}{*}{\centering\arraybackslash $\pi$} & $Y \times L^{(1)}$ & 2.321 & 2.321 & 2.321 & 2.321\\
\hline
\end{tabular}
\end{minipage}
\begin{minipage}[t]{0.5\textwidth}

\begin{tabular}[t]{>{}ccccc>{}c}
\hline
\multicolumn{1}{c}{\textbf{ }} & \multicolumn{1}{c}{\textbf{ }} & \multicolumn{4}{c}{\textbf{Data Driven Simulation Scenario}} \\
\cmidrule{3-6}
\textbf{Model} & \textbf{Term} & \textbf{13} & \textbf{14} & \textbf{15} & \textbf{16}\\
\hline
 & (Intercept) & 0.939 & 0.939 & 0.939 & 0.939\\
\cmidrule{2-6}
 & $L^{(1)}$ & 0.074 & 0.074 & 0.074 & 0.074\\
\cmidrule{2-6}
 & $L^{(2)}$ & -0.018 & -0.018 & -0.018 & -0.018\\
\cmidrule{2-6}
 & $L^{(3)}$ & -0.033 & -0.033 & -0.033 & -0.033\\
\cmidrule{2-6}
 & $(L^{(2)})^2$ & < 0.001 & < 0.001 & < 0.001 & < 0.001\\
\cmidrule{2-6}
 & $L^{(1)} \times L^{(2)}$ & --- & --- & --- & ---\\
\cmidrule{2-6}
 & $L^{(1)} \times L^{(3)}$ & --- & --- & --- & ---\\
\cmidrule{2-6}
 & $A$ & 0.102 & 0.102 & 0.302 & 0.302\\
\cmidrule{2-6}
 & $Y$ & -0.696 & -0.696 & -0.696 & -0.696\\
\cmidrule{2-6}
 & $A\times Y$ & -0.050 & -0.050 & -0.500 & -0.500\\
\cmidrule{2-6}
 & $Y \times L^{(2)}$ & -0.005 & -0.005 & -0.005 & -0.005\\
\cmidrule{2-6}
\multirow{-12}{*}{\centering\arraybackslash $\lambda_1~\text{or}~\tilde\lambda_1$} & $\alpha$ & 1.000 & 1.000 & 1.000 & 1.000\\
\cmidrule{1-6}
 & (Intercept) & -1.386 & -1.386 & -1.386 & -1.386\\
\cmidrule{2-6}
 & $L^{(1)}$ & 0.070 & --- & --- & 0.070\\
\cmidrule{2-6}
 & $L^{(2)}$ & 0.040 & --- & --- & 0.040\\
\cmidrule{2-6}
 & $L^{(3)}$ & 0.050 & --- & --- & 0.050\\
\cmidrule{2-6}
 & $L^{(4)}$ & 0.050 & 0.075 & 0.075 & 0.050\\
\cmidrule{2-6}
 & $A$ & 0.030 & --- & --- & 0.030\\
\cmidrule{2-6}
 & $Y$ & 0.250 & --- & -5.000 & -3.750\\
\cmidrule{2-6}
 & $A\times Y$ & --- & --- & --- & ---\\
\cmidrule{2-6}
 & $A \times L^{(4)}$ & --- & --- & --- & ---\\
\cmidrule{2-6}
 & $Y \times L^{(2)}$ & 0.050 & --- & --- & 0.050\\
\cmidrule{2-6}
\multirow{-11}{*}{\centering\arraybackslash $\lambda_2~\text{or}~\tilde\lambda_2$} & $Y \times L^{(4)}$ & --- & --- & --- & ---\\
\hline
\end{tabular}
\end{minipage}

\label{tab:coefficients_sup3}
\caption{Coefficients ($\beta$) and covariates ($X$) used to generated simulated datasets for simulation scenarios 13-16. All four scenarios generated datasets using the factorization from Equation (1). --- denotes the given covariate was not including in the corresponding model.}
\end{sidewaystable}

\begin{sidewaystable}
    \footnotesize
\begin{minipage}[t]{0.5\textwidth}
\centering

\begin{tabular}[t]{>{}cccc>{}c}
\hline
\multicolumn{1}{c}{\textbf{ }} & \multicolumn{1}{c}{\textbf{ }} & \multicolumn{3}{c}{\textbf{Data Driven Simulation Scenario}} \\
\cmidrule{3-5}
\textbf{Model} & \textbf{Term} & \textbf{17} & \textbf{18} & \textbf{19}\\
\hline
 & (Intercept) & -0.624 & -0.586 & -0.586\\
\cmidrule{2-5}
 & $L^{(1)}$ & 0.308 & 0.311 & 0.311\\
\cmidrule{2-5}
 & $L^{(2)}$ & -0.046 & -0.046 & -0.046\\
\cmidrule{2-5}
 & $L^{(3)}$ & 0.385 & 0.384 & 0.384\\
\cmidrule{2-5}
 & $L^{(4)}$ & --- & -0.015 & -0.035\\
\cmidrule{2-5}
 & $(L^{(2)})^2$ & 0.001 & 0.001 & 0.001\\
\cmidrule{2-5}
 & $L^{(1)} \times L^{(3)}$ & 0.400 & 0.100 & 0.100\\
\cmidrule{2-5}
\multirow{-8}{*}{\centering\arraybackslash $\eta~\text{or}~\tilde\eta$} & $L^{(3)} \times L^{(4)}$ & --- & --- & -0.020\\
\cmidrule{1-5}
 & (Intercept) & -0.207 & -0.200 & -0.200\\
\cmidrule{2-5}
 & $L^{(1)}$ & 0.031 & 0.032 & 0.032\\
\cmidrule{2-5}
 & $L^{(2)}$ & -0.002 & -0.002 & -0.002\\
\cmidrule{2-5}
 & $L^{(3)}$ & 0.023 & 0.022 & 0.022\\
\cmidrule{2-5}
 & $L^{(4)}$ & --- & -0.002 & -0.004\\
\cmidrule{2-5}
 & $L^{(5)}$ & --- & --- & ---\\
\cmidrule{2-5}
 & $L^{(1)} \times L^{(3)}$ & -0.305 & -0.305 & -0.305\\
\cmidrule{2-5}
 & $L^{(4)} \times L^{(5)}$ & --- & --- & ---\\
\cmidrule{2-5}
 & $A$ & 0.045 & 0.047 & 0.047\\
\cmidrule{2-5}
 & $A \times L^{(1)}$ & 0.313 & 0.313 & 0.313\\
\cmidrule{2-5}
 & $A \times L^{(2)}$ & -0.001 & -0.001 & 0.005\\
\cmidrule{2-5}
 & $A \times L^{(3)}$ & 0.080 & 0.081 & 0.081\\
\cmidrule{2-5}
 & $A \times L^{(4)}$ & --- & < 0.001 & -0.002\\
\cmidrule{2-5}
 & $A \times L^{(5)}$ & --- & --- & ---\\
\cmidrule{2-5}
\multirow{-15}{*}{\centering\arraybackslash $\mu~\text{or}~\tilde\mu$} & $\sigma$ & 0.109 & 0.109 & 0.109\\
\cmidrule{1-5}
 & (Intercept) & 1.824 & 1.824 & 1.824\\
\cmidrule{2-5}
 & $L^{(1)}$ & 0.087 & 0.087 & 0.087\\
\cmidrule{2-5}
 & $L^{(2)}$ & 0.010 & 0.010 & 0.010\\
\cmidrule{2-5}
 & $L^{(3)}$ & -2.662 & -2.662 & -2.662\\
\cmidrule{2-5}
 & $L^{(2)} \times L^{(3)}$ & -0.149 & -0.149 & -0.149\\
\cmidrule{2-5}
 & $A$ & 2.922 & 2.922 & 2.922\\
\cmidrule{2-5}
 & $Y$ & 2.180 & 2.180 & 2.180\\
\cmidrule{2-5}
 & $A\times Y$ & 3.043 & 3.043 & 3.043\\
\cmidrule{2-5}
 & $A \times L^{(2)}$ & 0.159 & 0.159 & 0.159\\
\cmidrule{2-5}
\multirow{-10}{*}{\centering\arraybackslash $\pi$} & $Y \times L^{(1)}$ & 2.321 & 2.321 & 2.321\\
\hline
\end{tabular}
\end{minipage}
\begin{minipage}[t]{0.5\textwidth}

\begin{tabular}[t]{>{}cccc>{}c}
\hline
\multicolumn{1}{c}{\textbf{ }} & \multicolumn{1}{c}{\textbf{ }} & \multicolumn{3}{c}{\textbf{Data Driven Simulation Scenario}} \\
\cmidrule{3-5}
\textbf{Model} & \textbf{Term} & \textbf{17} & \textbf{18} & \textbf{19}\\
\hline
 & (Intercept) & 0.939 & 0.900 & 0.900\\
\cmidrule{2-5}
 & $L^{(1)}$ & 0.074 & 0.272 & 0.272\\
\cmidrule{2-5}
 & $L^{(2)}$ & -0.018 & 0.001 & 0.001\\
\cmidrule{2-5}
 & $L^{(3)}$ & -0.033 & -0.035 & -0.035\\
\cmidrule{2-5}
 & $(L^{(2)})^2$ & < 0.001 & --- & ---\\
\cmidrule{2-5}
 & $L^{(1)} \times L^{(2)}$ & --- & --- & 0.001\\
\cmidrule{2-5}
 & $L^{(1)} \times L^{(3)}$ & --- & 0.100 & 0.100\\
\cmidrule{2-5}
 & $A$ & 0.302 & --- & ---\\
\cmidrule{2-5}
 & $Y$ & -0.696 & --- & ---\\
\cmidrule{2-5}
 & $A\times Y$ & -0.500 & --- & ---\\
\cmidrule{2-5}
 & $Y \times L^{(2)}$ & -0.005 & --- & ---\\
\cmidrule{2-5}
\multirow{-12}{*}{\centering\arraybackslash $\lambda_1~\text{or}~\tilde\lambda_1$} & $\alpha$ & 1.000 & 3.616 & 3.616\\
\cmidrule{1-5}
 & (Intercept) & -1.386 & --- & ---\\
\cmidrule{2-5}
 & $L^{(1)}$ & 0.070 & --- & ---\\
\cmidrule{2-5}
 & $L^{(2)}$ & 0.040 & --- & ---\\
\cmidrule{2-5}
 & $L^{(3)}$ & 0.050 & --- & ---\\
\cmidrule{2-5}
 & $L^{(4)}$ & 0.050 & --- & ---\\
\cmidrule{2-5}
 & $A$ & 0.030 & --- & ---\\
\cmidrule{2-5}
 & $Y$ & -4.000 & --- & ---\\
\cmidrule{2-5}
 & $A\times Y$ & 2.000 & --- & ---\\
\cmidrule{2-5}
 & $A \times L^{(4)}$ & --- & --- & ---\\
\cmidrule{2-5}
 & $Y \times L^{(2)}$ & 0.050 & --- & ---\\
\cmidrule{2-5}
\multirow{-11}{*}{\centering\arraybackslash $\lambda_2~\text{or}~\tilde\lambda_2$} & $Y \times L^{(4)}$ & --- & --- & ---\\
\hline
\end{tabular}
\end{minipage}

\caption{Coefficients ($\beta$) and covariates ($X$) used to generated simulated datasets for simulation scenarios 17-18. Scenario 17 generated datasets using the factorization from Equation (1), while scenarios 18-19 generated datasets from the alternative data factorization. --- denotes the given covariate was not including in the corresponding model. Note that scenarios 18-19 only generated a single partially missing confounder and as such no $\lambda_2$ model was needed.}\label{tab:coefficients_sup4}
\end{sidewaystable}

\subsection{Simulation Results}\label{sec:sim_results}
Simulation results for scenarios \#5-19 are presented in Tables \ref{table:results_5} -  \ref{table:results_19}. Key takeaways from these results do not differ from the summary presented in the Results section in the body of the main paper. Nevertheless, such results reinforce the notion that our findings and interpretation are robust to a wide range of scenarios.
\begin{table}
\centering\footnotesize
\begin{tabularx}{\textwidth}{c@{}c@{}c@{}c@{}c@{}cc@{}c}
\hline
\multicolumn{8}{c}{\textbf{Data Driven Simulation Scenario 5}} \\
\cmidrule{1-8}
\multicolumn{2}{c}{\textbf{Model}} & \textbf{Interactions} & \textbf{Imputation} & \textbf{\% Bias} & \textbf{\% M-Bias} & \textbf{SE} & \textbf{RU}\\
\hline
 & IF &  & --- & -2 & 1 & 0.021 & 1.000\\
\cmidrule{2-2}
\cmidrule{4-8}
\multirow{-2}{*}{\centering\arraybackslash \shortstack{CCMAR-\\based}} & IWOR & \multirow{-2}{*}{\centering\arraybackslash ---} & --- & 1 & -4 & 0.014 & 0.687\\
\cmidrule{1-8}
 &  &  & OLS/LR & 24 & 24 & 0.005 & 0.251\\
\cmidrule{4-8}
 &  &  & OLS/LR w/ Interactions & 23 & 23 & 0.006 & 0.277\\
\cmidrule{4-8}
 &  &  & True DGP & 26 & 26 & 0.005 & 0.254\\
\cmidrule{4-8}
 &  & \multirow{-4}{*}{\centering\arraybackslash None} & --- & -13 & -13 & 0.005 & 0.266\\
\end{tabularx}\caption{Data Driven Simulation Scenario \#5: Replication of simulation from Levis \etal\cite{levis2022} paper: Factorization in Equation (1), 1 partially missing confounder.}
\label{table:results_5}
\end{table}

\begin{table}
\centering\footnotesize
\begin{tabularx}{\textwidth}{c@{}c@{}c@{}c@{}c@{}cc@{}c}
\hline
\multicolumn{8}{c}{\textbf{Data Driven Simulation Scenario 6}} \\
\cmidrule{1-8}
\multicolumn{2}{c}{\textbf{Model}} & \textbf{Interactions} & \textbf{Imputation} & \textbf{\% Bias} & \textbf{\% M-Bias} & \textbf{SE} & \textbf{RU}\\
\hline
 & IF &  & --- & 0 & 0 & 0.005 & 1.000\\
\cmidrule{2-2}
\cmidrule{4-8}
\multirow{-2}{*}{\centering\arraybackslash \shortstack{CCMAR-\\based}} & IWOR & \multirow{-2}{*}{\centering\arraybackslash ---} & --- & -1 & -5 & 0.023 & 4.394\\
\cmidrule{1-8}
 &  &  & OLS/LR & 7 & 7 & 0.005 & 0.922\\
\cmidrule{4-8}
 &  &  & OLS/LR w/ Interactions & 7 & 7 & 0.005 & 0.929\\
\cmidrule{4-8}
 &  &  & True DGP & 7 & 7 & 0.005 & 0.922\\
\cmidrule{4-8}
 &  & \multirow{-4}{*}{\centering\arraybackslash None} & --- & -22 & -22 & 0.005 & 1.017\\
\end{tabularx}\caption{Data Driven Simulation Scenario \#6: Factorization in Equation (1), 1 partially missing confounder, dampen some of the effects amplified in treatment model and imputation model in scenario \#5 to correct near-positivity violations.}
\label{table:results_6}
\end{table}

\begin{table}
\centering\footnotesize
\begin{tabularx}{\textwidth}{c@{}c@{}c@{}c@{}c@{}cc@{}c}
\hline
\multicolumn{8}{c}{\textbf{Data Driven Simulation Scenario 7}} \\
\cmidrule{1-8}
\multicolumn{2}{c}{\textbf{Model}} & \textbf{Interactions} & \textbf{Imputation} & \textbf{\% Bias} & \textbf{\% M-Bias} & \textbf{SE} & \textbf{RU}\\
\hline
 & IF &  & --- & 0 & 0 & 0.005 & 1.000\\
\cmidrule{2-2}
\cmidrule{4-8}
\multirow{-2}{*}{\centering\arraybackslash \shortstack{CCMAR-\\based}} & IWOR & \multirow{-2}{*}{\centering\arraybackslash ---} & --- & -1 & -5 & 0.020 & 4.098\\
\cmidrule{1-8}
 &  &  & OLS/LR & 5 & 5 & 0.005 & 0.987\\
\cmidrule{4-8}
 &  &  & OLS/LR w/ Interactions & 5 & 5 & 0.005 & 0.989\\
\cmidrule{4-8}
 &  &  & True DGP & 5 & 5 & 0.005 & 0.987\\
\cmidrule{4-8}
 &  & \multirow{-4}{*}{\centering\arraybackslash None} & --- & -22 & -22 & 0.005 & 1.103\\
\end{tabularx}\caption{Data Driven Simulation Scenario \#7: Factorization in Equation (1), 1 partially missing confounder, amplify interactions in treatment model.}
\label{table:results_7}
\end{table}

\begin{table}
\centering\footnotesize
\begin{tabularx}{\textwidth}{c@{}c@{}c@{}c@{}c@{}cc@{}c}
\hline
\multicolumn{8}{c}{\textbf{Data Driven Simulation Scenario 8}} \\
\cmidrule{1-8}
\multicolumn{2}{c}{\textbf{Model}} & \textbf{Interactions} & \textbf{Imputation} & \textbf{\% Bias} & \textbf{\% M-Bias} & \textbf{SE} & \textbf{RU}\\
\hline
 & IF &  & --- & 0 & 0 & 0.005 & 1.000\\
\cmidrule{2-2}
\cmidrule{4-8}
\multirow{-2}{*}{\centering\arraybackslash \shortstack{CCMAR-\\based}} & IWOR & \multirow{-2}{*}{\centering\arraybackslash ---} & --- & -1 & -5 & 0.020 & 4.053\\
\cmidrule{1-8}
 &  &  & OLS/LR & 5 & 5 & 0.005 & 0.982\\
\cmidrule{4-8}
 &  &  & OLS/LR w/ Interactions & 5 & 5 & 0.005 & 0.987\\
\cmidrule{4-8}
 &  &  & True DGP & 5 & 5 & 0.005 & 0.982\\
\cmidrule{4-8}
 &  & \multirow{-4}{*}{\centering\arraybackslash None} & --- & -22 & -22 & 0.005 & 1.098\\
\end{tabularx}\caption{Data Driven Simulation Scenario \#8: Factorization in Equation (1), 1 partially missing confounder, introduce non-linearities in imputation model.}
\label{table:results_8}
\end{table}

\begin{table}
\centering\footnotesize
\begin{tabularx}{\textwidth}{c@{}c@{}c@{}c@{}c@{}cc@{}c}
\hline
\multicolumn{8}{c}{\textbf{Data Driven Simulation Scenario 9}} \\
\cmidrule{1-8}
\multicolumn{2}{c}{\textbf{Model}} & \textbf{Interactions} & \textbf{Imputation} & \textbf{\% Bias} & \textbf{\% M-Bias} & \textbf{SE} & \textbf{RU}\\
\hline
 & IF &  & --- & 0 & 0 & 0.008 & 1.000\\
\cmidrule{2-2}
\cmidrule{4-8}
\multirow{-2}{*}{\centering\arraybackslash \shortstack{CCMAR-\\based}} & IWOR & \multirow{-2}{*}{\centering\arraybackslash ---} & --- & -1 & -5 & 0.023 & 3.028\\
\cmidrule{1-8}
 &  &  & OLS/LR & 7 & 7 & 0.005 & 0.638\\
\cmidrule{4-8}
 &  &  & OLS/LR w/ Interactions & 7 & 7 & 0.005 & 0.643\\
\cmidrule{4-8}
 &  &  & True DGP & 7 & 7 & 0.005 & 0.638\\
\cmidrule{4-8}
 &  & \multirow{-4}{*}{\centering\arraybackslash None} & --- & -22 & -22 & 0.005 & 0.703\\
\end{tabularx}\caption{Data Driven Simulation Scenario \#9: Factorization in Equation (1), 2 partially missing confounders, extend scenario \#6 to have a second partially missing confounder.}
\label{table:results_9}
\end{table}

\begin{table}
\centering\footnotesize
\begin{tabularx}{\textwidth}{c@{}c@{}c@{}c@{}c@{}cc@{}c}
\hline
\multicolumn{8}{c}{\textbf{Data Driven Simulation Scenario 10}} \\
\cmidrule{1-8}
\multicolumn{2}{c}{\textbf{Model}} & \textbf{Interactions} & \textbf{Imputation} & \textbf{\% Bias} & \textbf{\% M-Bias} & \textbf{SE} & \textbf{RU}\\
\hline
 & IF &  & --- & 0 & 0 & 0.006 & 1.000\\
\cmidrule{2-2}
\cmidrule{4-8}
\multirow{-2}{*}{\centering\arraybackslash \shortstack{CCMAR-\\based}} & IWOR & \multirow{-2}{*}{\centering\arraybackslash ---} & --- & -1 & -5 & 0.020 & 3.126\\
\cmidrule{1-8}
 &  &  & OLS/LR & 5 & 5 & 0.005 & 0.760\\
\cmidrule{4-8}
 &  &  & OLS/LR w/ Interactions & 5 & 5 & 0.005 & 0.761\\
\cmidrule{4-8}
 &  &  & True DGP & 5 & 5 & 0.005 & 0.760\\
\cmidrule{4-8}
 &  & \multirow{-4}{*}{\centering\arraybackslash None} & --- & -22 & -22 & 0.005 & 0.848\\
\end{tabularx}\caption{Data Driven Simulation Scenario \#10: Factorization in Equation (1), 2 partially missing confounders, extend scenario \#7 to have a second partially missing confounder.}
\label{table:results_10}
\end{table}

\begin{table}
\centering\footnotesize
\begin{tabularx}{\textwidth}{c@{}c@{}c@{}c@{}c@{}cc@{}c}
\hline
\multicolumn{8}{c}{\textbf{Data Driven Simulation Scenario 11}} \\
\cmidrule{1-8}
\multicolumn{2}{c}{\textbf{Model}} & \textbf{Interactions} & \textbf{Imputation} & \textbf{\% Bias} & \textbf{\% M-Bias} & \textbf{SE} & \textbf{RU}\\
\hline
 & IF &  & --- & 0 & 0 & 0.007 & 1.000\\
\cmidrule{2-2}
\cmidrule{4-8}
\multirow{-2}{*}{\centering\arraybackslash \shortstack{CCMAR-\\based}} & IWOR & \multirow{-2}{*}{\centering\arraybackslash ---} & --- & -1 & -6 & 0.032 & 4.796\\
\cmidrule{1-8}
 &  &  & OLS/LR & 15 & 15 & 0.006 & 0.881\\
\cmidrule{4-8}
 &  &  & OLS/LR w/ Interactions & 12 & 12 & 0.006 & 0.892\\
\cmidrule{4-8}
 &  &  & True DGP & 14 & 14 & 0.006 & 0.881\\
\cmidrule{4-8}
 &  & \multirow{-4}{*}{\centering\arraybackslash None} & --- & -21 & -21 & 0.006 & 0.919\\
\end{tabularx}\caption{Data Driven Simulation Scenario \#11: Factorization in Equation (1), 2 partially missing confounders, extend scenario \#1 to have a second partially missing confounder (baseline simulation in main paper with a second missing confounder).}
\label{table:results_11}
\end{table}

\begin{table}
\centering\footnotesize
\begin{tabularx}{\textwidth}{c@{}c@{}c@{}c@{}c@{}cc@{}c}
\hline
\multicolumn{8}{c}{\textbf{Data Driven Simulation Scenario 12}} \\
\cmidrule{1-8}
\multicolumn{2}{c}{\textbf{Model}} & \textbf{Interactions} & \textbf{Imputation} & \textbf{\% Bias} & \textbf{\% M-Bias} & \textbf{SE} & \textbf{RU}\\
\hline
 & IF &  & --- & 0 & 0 & 0.006 & 1.000\\
\cmidrule{2-2}
\cmidrule{4-8}
\multirow{-2}{*}{\centering\arraybackslash \shortstack{CCMAR-\\based}} & IWOR & \multirow{-2}{*}{\centering\arraybackslash ---} & --- & -1 & -5 & 0.020 & 3.127\\
\cmidrule{1-8}
 &  &  & OLS/LR & 5 & 5 & 0.005 & 0.764\\
\cmidrule{4-8}
 &  &  & OLS/LR w/ Interactions & 5 & 5 & 0.005 & 0.768\\
\cmidrule{4-8}
 &  &  & True DGP & 5 & 5 & 0.005 & 0.764\\
\cmidrule{4-8}
 &  & \multirow{-4}{*}{\centering\arraybackslash None} & --- & -22 & -22 & 0.005 & 0.853\\
\end{tabularx}\caption{Data Driven Simulation Scenario \#12: Factorization in Equation (1), 2 partially missing confounder, extend scenario \#8 to have a second partially missing confounder.}
\label{table:results_12}
\end{table}

\begin{table}
\centering\footnotesize
\begin{tabularx}{\textwidth}{c@{}c@{}c@{}c@{}c@{}cc@{}c}
\hline
\multicolumn{8}{c}{\textbf{Data Driven Simulation Scenario 13}} \\
\cmidrule{1-8}
\multicolumn{2}{c}{\textbf{Model}} & \textbf{Interactions} & \textbf{Imputation} & \textbf{\% Bias} & \textbf{\% M-Bias} & \textbf{SE} & \textbf{RU}\\
\hline
 & IF &  & --- & 0 & 0 & 0.006 & 1.000\\
\cmidrule{2-2}
\cmidrule{4-8}
\multirow{-2}{*}{\centering\arraybackslash \shortstack{CCMAR-\\based}} & IWOR & \multirow{-2}{*}{\centering\arraybackslash ---} & --- & -1 & -5 & 0.020 & 3.316\\
\cmidrule{1-8}
 &  &  & OLS/LR & 5 & 5 & 0.005 & 0.801\\
\cmidrule{4-8}
 &  &  & OLS/LR w/ Interactions & 5 & 5 & 0.005 & 0.801\\
\cmidrule{4-8}
 &  &  & True DGP & 5 & 5 & 0.005 & 0.800\\
\cmidrule{4-8}
 &  & \multirow{-4}{*}{\centering\arraybackslash None} & --- & -23 & -23 & 0.005 & 0.898\\
\end{tabularx}\caption{Data Driven Simulation Scenario \#13: Factorization in Equation (1), 2 partially missing confounder, extend scenario \#2 to have a second partially missing confounder.}
\label{table:results_13}
\end{table}

\begin{table}
\centering\footnotesize
\begin{tabularx}{\textwidth}{c@{}c@{}c@{}c@{}c@{}cc@{}c}
\hline
\multicolumn{8}{c}{\textbf{Data Driven Simulation Scenario 14}} \\
\cmidrule{1-8}
\multicolumn{2}{c}{\textbf{Model}} & \textbf{Interactions} & \textbf{Imputation} & \textbf{\% Bias} & \textbf{\% M-Bias} & \textbf{SE} & \textbf{RU}\\
\hline
 & IF &  & --- & 0 & 0 & 0.008 & 1.000\\
\cmidrule{2-2}
\cmidrule{4-8}
\multirow{-2}{*}{\centering\arraybackslash \shortstack{CCMAR-\\based}} & IWOR & \multirow{-2}{*}{\centering\arraybackslash ---} & --- & -1 & -5 & 0.020 & 2.653\\
\cmidrule{1-8}
 &  &  & OLS/LR & 5 & 5 & 0.005 & 0.637\\
\cmidrule{4-8}
 &  &  & OLS/LR w/ Interactions & 5 & 5 & 0.005 & 0.636\\
\cmidrule{4-8}
 &  &  & True DGP & 5 & 5 & 0.005 & 0.636\\
\cmidrule{4-8}
 &  & \multirow{-4}{*}{\centering\arraybackslash None} & --- & -23 & -23 & 0.005 & 0.715\\
\end{tabularx}\caption{Data Driven Simulation Scenario \#14: Factorization in Equation (1), 2 partially missing confounders, additional non-linearities and interactions in imputation model for first missing confounder, induce skew in imputation model for first missing confounder, imputation model for second missing confounder depends only on $L_{p1}$ with amplified relationship to emphasize importance of modeling $\lambda_1(\ell_{p1} ~|~ L_c, A, Y, S = 1)$ correctly.}
\label{table:results_14}
\end{table}

\begin{table}
\centering\footnotesize
\begin{tabularx}{\textwidth}{c@{}c@{}c@{}c@{}c@{}cc@{}c}
\hline
\multicolumn{8}{c}{\textbf{Data Driven Simulation Scenario 15}} \\
\cmidrule{1-8}
\multicolumn{2}{c}{\textbf{Model}} & \textbf{Interactions} & \textbf{Imputation} & \textbf{\% Bias} & \textbf{\% M-Bias} & \textbf{SE} & \textbf{RU}\\
\hline
 & IF &  & --- & 0 & 0 & 0.007 & 1.000\\
\cmidrule{2-2}
\cmidrule{4-8}
\multirow{-2}{*}{\centering\arraybackslash \shortstack{CCMAR-\\based}} & IWOR & \multirow{-2}{*}{\centering\arraybackslash ---} & --- & 0 & -4 & 0.019 & 2.861\\
\cmidrule{1-8}
 &  &  & OLS/LR & 5 & 5 & 0.005 & 0.727\\
\cmidrule{4-8}
 &  &  & OLS/LR w/ Interactions & 4 & 4 & 0.005 & 0.715\\
\cmidrule{4-8}
 &  &  & True DGP & 4 & 4 & 0.005 & 0.707\\
\cmidrule{4-8}
 &  & \multirow{-4}{*}{\centering\arraybackslash None} & --- & -21 & -21 & 0.005 & 0.794\\
\end{tabularx}\caption{Data Driven Simulation Scenario \#15: Factorization in Equation (1), 2 partially missing confounders, extend scenario \#14 such that imputation model for second missing confounder depends only on both $L_{p1}$ and $Y$.}
\label{table:results_15}
\end{table}

\begin{table}
\centering\footnotesize
\begin{tabularx}{\textwidth}{c@{}c@{}c@{}c@{}c@{}cc@{}c}
\hline
\multicolumn{8}{c}{\textbf{Data Driven Simulation Scenario 16}} \\
\cmidrule{1-8}
\multicolumn{2}{c}{\textbf{Model}} & \textbf{Interactions} & \textbf{Imputation} & \textbf{\% Bias} & \textbf{\% M-Bias} & \textbf{SE} & \textbf{RU}\\
\hline
 & IF &  & --- & 0 & 0 & 0.006 & 1.000\\
\cmidrule{2-2}
\cmidrule{4-8}
\multirow{-2}{*}{\centering\arraybackslash \shortstack{CCMAR-\\based}} & IWOR & \multirow{-2}{*}{\centering\arraybackslash ---} & --- & -1 & -5 & 0.020 & 3.345\\
\cmidrule{1-8}
 &  &  & OLS/LR & 6 & 6 & 0.005 & 0.819\\
\cmidrule{4-8}
 &  &  & OLS/LR w/ Interactions & 5 & 5 & 0.005 & 0.814\\
\cmidrule{4-8}
 &  &  & True DGP & 5 & 5 & 0.005 & 0.814\\
\cmidrule{4-8}
 &  & \multirow{-4}{*}{\centering\arraybackslash None} & --- & -21 & -21 & 0.005 & 0.902\\
\end{tabularx}\caption{Data Driven Simulation Scenario \#16: Factorization in Equation (1),, 2 partially missing confounders amplify effect of outcome $Y$ in imputation model for second missing confounder.}
\label{table:results_16}
\end{table}

\begin{table}
\centering\footnotesize
\begin{tabularx}{\textwidth}{c@{}c@{}c@{}c@{}c@{}cc@{}c}
\hline
\multicolumn{8}{c}{\textbf{Data Driven Simulation Scenario 17}} \\
\cmidrule{1-8}
\multicolumn{2}{c}{\textbf{Model}} & \textbf{Interactions} & \textbf{Imputation} & \textbf{\% Bias} & \textbf{\% M-Bias} & \textbf{SE} & \textbf{RU}\\
\hline
 & IF &  & --- & 0 & 0 & 0.006 & 1.000\\
\cmidrule{2-2}
\cmidrule{4-8}
\multirow{-2}{*}{\centering\arraybackslash \shortstack{CCMAR-\\based}} & IWOR & \multirow{-2}{*}{\centering\arraybackslash ---} & --- & -1 & -5 & 0.020 & 3.388\\
\cmidrule{1-8}
 &  &  & OLS/LR & 5 & 5 & 0.005 & 0.840\\
\cmidrule{4-8}
 &  &  & OLS/LR w/ Interactions & 4 & 4 & 0.005 & 0.837\\
\cmidrule{4-8}
 &  &  & True DGP & 4 & 4 & 0.005 & 0.833\\
\cmidrule{4-8}
 &  & \multirow{-4}{*}{\centering\arraybackslash None} & --- & -22 & -22 & 0.006 & 0.919\\
\end{tabularx}\caption{Data Driven Simulation Scenario \#17: Factorization in Equation (1), 2 partially missing confounders, introduce interaction between treatment $A$ and outcome $Y$ in imputation model for second missing confounde.}
\label{table:results_17}
\end{table}

\begin{table}
\centering\footnotesize
\begin{tabularx}{\textwidth}{c@{}c@{}c@{}c@{}c@{}cc@{}c}
\hline
\multicolumn{8}{c}{\textbf{Data Driven Simulation Scenario 18}} \\
\cmidrule{1-8}
\multicolumn{2}{c}{\textbf{Model}} & \textbf{Interactions} & \textbf{Imputation} & \textbf{\% Bias} & \textbf{\% M-Bias} & \textbf{SE} & \textbf{RU}\\
\hline
 & IF &  & --- & 0 & 0 & 0.004 & 1.000\\
\cmidrule{2-2}
\cmidrule{4-8}
\multirow{-2}{*}{\centering\arraybackslash \shortstack{CCMAR-\\based}} & IWOR & \multirow{-2}{*}{\centering\arraybackslash ---} & --- & -1 & -6 & 0.020 & 4.712\\
\cmidrule{1-8}
 &  &  & OLS/LR & 7 & 7 & 0.005 & 1.160\\
\cmidrule{4-8}
 &  &  & OLS/LR w/ Interactions & 7 & 7 & 0.005 & 1.160\\
\cmidrule{4-8}
 &  &  & True DGP & 7 & 7 & 0.005 & 1.159\\
\cmidrule{4-8}
 &  & \multirow{-4}{*}{\centering\arraybackslash None} & --- & -23 & -23 & 0.005 & 1.265\\
\end{tabularx}\caption{Data Driven Simulation Scenario \#18: Alternative factorization, 1 partially missing confounder.}
\label{table:results_18}
\end{table}

\begin{table}
\centering\footnotesize
\begin{tabularx}{\textwidth}{c@{}c@{}c@{}c@{}c@{}cc@{}c}
\hline
\multicolumn{8}{c}{\textbf{Data Driven Simulation Scenario 19}} \\
\cmidrule{1-8}
\multicolumn{2}{c}{\textbf{Model}} & \textbf{Interactions} & \textbf{Imputation} & \textbf{\% Bias} & \textbf{\% M-Bias} & \textbf{SE} & \textbf{RU}\\
\hline
 & IF &  & --- & 0 & 0 & 0.004 & 1.000\\
\cmidrule{2-2}
\cmidrule{4-8}
\multirow{-2}{*}{\centering\arraybackslash \shortstack{CCMAR-\\based}} & IWOR & \multirow{-2}{*}{\centering\arraybackslash ---} & --- & -1 & -5 & 0.030 & 6.629\\
\cmidrule{1-8}
 &  &  & OLS/LR & 3 & 3 & 0.005 & 1.144\\
\cmidrule{4-8}
 &  &  & OLS/LR w/ Interactions & 3 & 3 & 0.005 & 1.146\\
\cmidrule{4-8}
 &  &  & True DGP & 3 & 3 & 0.005 & 1.143\\
\cmidrule{4-8}
 &  & \multirow{-4}{*}{\centering\arraybackslash None} & --- & -15 & -15 & 0.006 & 1.234\\
\end{tabularx}\caption{Data Driven Simulation Scenario \#19: Alternative factorization, 1 partially missing confounder, introduce interactions between $L_c$ and $L_p$ in treatment model $\tilde\eta(L_c, L_p a)$.}
\label{table:results_19}
\end{table}

\clearpage
\section{Definition of Complete-Case Missing At Random (CCMAR) Estimators}
Definitions of the $\tilde{\chi}_a$ and $\hat{\chi}_a$, the complete-case missing at random (CCMAR) estimators proposed by Levis and colleagues\cite{levis2022} are available in Sections \ref{sec:iwor} and \ref{sec:if}, respectively. In order to provide these definitions, we must first define some additional quantities.

$$
\begin{aligned}
\xi(L_c, a; L_p) &= \frac{\beta(L_c, a; L_p)}{\gamma(L_c, a; L_p)} =\ \frac{\int_\mathcal{Y} y \lambda(L_p \mid L_c, a, y, S)\, d \mu(y \mid L_c, a) }{\int_\mathcal{Y} \lambda(L_p \mid L_c, a, y, S)\, d \mu(y \mid L_c, a)} \\
\hat \tau(L_c; L_p) &=\ \sum_{a'=0}^1\hat \eta(L_c, a')\hat \gamma(L_c, a'; L_p) \\
b_{a1}(L_c, A, Y) &= \mathbb{E}_{P_{o^\prime}}\left[\frac{\beta(L_c, a; L_p)}{\gamma(L_c, a; L_p)} 
    \bigg| \, L_c, A, Y, S=1\right] \\
b_{a2}(L_c, Y) &= \mathbb{E}_{P_{o^\prime}}\left[\frac{\tau(L_c; L_p)}{\gamma(L_c, a; L_p)}
    \left(Y - \frac{\beta(L_c, a; L_p)}{\gamma(L_c, a; L_p)}\right)  \bigg
    \vert L_c, A = a, Y, S=1 \right]
\end{aligned}
$$
\\~\\These expressions are estimated from the nuisance functions outlined in Table 1 of the main paper as follows:

$$
\begin{aligned}
\hat{\xi}(L_c, a, L_p) &= \frac{\hat{\gamma}(L_c, a, L_p)}{\hat{\beta}(L_c, a, L_p)} \\
\hat{\gamma}(L_c, a, L_p) &= \int_{\mathcal{Y}}  \hat{\lambda}(L_p \mid L_c, a, y) d \hat{\mu}(y \mid L_c, a)\\
\hat{\beta}(L_c, a,L_p) &= \int_{\mathcal{Y}} y \hat{\lambda}(L_p\mid L_c, a, y) d\hat{\mu}(y \mid L_c, a) \\
\hat \tau(L_c; L_p) &=\ \sum_{a'=0}^1\hat \eta(L_c, a')\hat \gamma(L_c, a'; L_p) \\
\hat{b}_{a1}(L_c, A, Y) &= \int_{\mathcal{L}_p}\frac{\hat{\beta}(L_c, a; \ell_p)}{\hat{\gamma}(L_c, a; \ell_p)}\hat{\lambda}(\ell_p \mid L_c, A, Y) \, d\nu(\ell_p) \\
\hat{b}_{a2}(L_c, Y) &=  \int_{\mathcal{L}_p}\frac{\hat{\tau}(L_c; \ell_p)}{\hat{\gamma}(L_c, a; \ell_p)} \bigg(Y - \frac{\hat{\beta}(L_c, a; \ell_p)}{\hat{\gamma}(L_c, a; \ell_p)}\bigg) \hat{\lambda}(\ell_p \mid  L_c, a, Y) d\nu(\ell_p) \\
\end{aligned}
$$
\\~\\In the above, $\nu$ is the dominating measure for the density $\lambda$. Practically, these integrals become sums when $Y$ is discrete, but otherwise will typically require approximation via stochastic (e.g., Monte Carlo) or deterministic (e.g., Gauss-Hermite quadrature) methods. 

\subsection{Inverse Weighted Outcome Regression (IWOR) Estimator}\label{sec:iwor}
The complete-case missing at random inverse weighted outcome regression (IWOR) estimator of $\mathbb{E}[Y(a)]$ proposed by Levis and colleagues\cite{levis2022} is defined as:

$$
\tilde{\chi}_a\ =\ \frac{1}{n} \sum\limits_{i=1}^n \frac{S_i}{\hat{\pi}(L_{c,i}, A_i, Y_i)} \hat{\xi}(L_{c,i}, a; L_{p,i})
$$

\subsection{Influence Function-Based Estimator}\label{sec:if}

The complete-case missing at random influence function-based (IF) estimator of $\mathbb{E}[Y(a)]$ proposed by Levis and colleagues\cite{levis2022} is defined as:

\begin{align*}
  \hat{\chi}_a
  &= \frac{1}{n}\sum_{i=1}^n \biggr[\hat{b}_{a1}(L_{c, i}, A_i, Y_i) +
    \frac{\mathds{1}(A_i = a)}{\hat{\eta}(L_{c, i}, a)}
    \hat{b}_{a2}(L_{c, i}, Y_i) \\
  & \quad \quad + \frac{S}{\hat{\pi}(L_{c, i}, A_i,
    Y_i)} \bigg\{\frac{\hat{\beta}(L_{c, i}, a; L_{p, i})}
    {\hat{\gamma}(L_{c, i}, a; L_{p, i})} -  \hat{b}_{a1}(L_{c, i}, A_i, Y_i) \\
  & \quad  \quad + \frac{\mathds{1}(A_i = a)}{\hat{\eta}(L_{c, i}, a)}\left(
    \frac{\hat{\tau}(L_{c, i}; L_{p, i})}{\hat{\gamma}(L_{c, i}, a; L_{p, i})}
    \left(Y_i - \frac{\hat{\beta}(L_{c, i}, a; L_{p, i})}
    {\hat{\gamma}(L_{c, i}, a; L_{p, i})}\right)  -
    \hat{b}_{a2}(L_{c, i}, Y_i) \right)\bigg\}\biggr]
\end{align*}

\section{Identification Assumptions}
Identification assumptions for $\mathbb{E}[Y(a)]$ under the setting considered in this work are as follows:

\begin{enumerate}
    \item Consistency $Y(A) = Y$
    \item No unmeasured confounding $A \indep Y(a)~|~L$ for all $a \in \mathcal{A}$
    \item Positivity $\epsilon < P_f(A = a~|~L)< 1 - \epsilon$ for some $\epsilon > 0$, for all $a \in \mathcal{A}$
    \item Complete-case missing at random $S \indep L_p~|~ L_c, A, Y$
    \item Complete-case positivity $\epsilon < P_{o'}(S = 1~|~L_c, A, Y)$ for some $\epsilon > 0$
\end{enumerate}

Assumptions 1-3 are standard causal inference assumptions for identification of $\mathbb{E}[Y(a)]$ in the absence of missing data. Assumptions 4-5 are missing data assumptions, discussed in great detail in $\S3.2$ of Levis \etal\cite{levis2022}. Proofs of identification under these assumptions are provided in the supplementary materials of Levis \etal In particular, identification of the ATE is given by

\begin{equation}\label{eqn:identification}
\mathbb{E}_{P_{o'}}\biggr[\frac{S}{\pi(L_c, A, Y)}\xi(L_c, a; L_p)\biggr]    
\end{equation}

\section{Computation of True Average Treatment Effect}

In order to compute bias in any simulation setting, a ground truth value of the ATE is necessary. When simulating data from factorization in Equation (1), we computed the true ATE directly from the identification result in Equation \ref{eqn:identification}, using the coefficient values from known nuisance models used to generate simulated data. Expectations are taken over the entire Kaiser Permanente population from which covariate vectors $L_c$ were sampled.

When simulating data from factorization (3), the identification result in  Equation \ref{eqn:identification} could not be applied directly, as the coefficient values for necessary nuisance models (e.g., those that would be induced for factorization (1) ) were unknown. Thus, to obtain the true ATE in these situations, a large ($n = 10,000,000$) was simulated under factorization (3), without any missing data. Upon fitting correctly specified models for each relevant nuisance function, we applied the $g$-formula to get a ``true'' ATE against which to bench mark estimators.

\bibliography{bibliography}